\documentclass[%
 reprint,
 amsmath,amssymb,
 aps,
 prl,
]{revtex4-1}
\usepackage[english]{babel}
\usepackage[colorlinks=true,linkcolor=blue,citecolor=blue,urlcolor=blue]{hyperref}
\usepackage{graphicx}
\usepackage{placeins}
\usepackage{dcolumn}
\usepackage{bm}

\usepackage{comment}

\newcommand{\ket}{\rangle}
\newcommand{\bra}{\langle}
\newcommand{\Ket}{\Big\rangle}

\newcommand{\Line}{\Big|}
\newcommand{\degree}{$^{\circ}$}

\newcommand{\sixJ}[6] {\left\{\begin{array}{ccc} #1 & #2 & #3 \\ #4 & #5 & #6 \end{array}\right \}}
\newcommand{\threeJ}[6]{\left(\begin{array}{ccc} #1 & #3 & #5 \\ #2 & #4 & #6 \end{array}\right )}

\usepackage{rotating}

\usepackage[dvipsnames]{xcolor}

\begin{document}

\title{Probing Fundamental Symmetries of Deformed Nuclei in Symmetric Top Molecules}

\author{Phelan Yu}
\email{phelanyu@caltech.edu}
\author{Nicholas R. Hutzler}
\email{hutzler@caltech.edu}
\affiliation{Division of Physics, Mathematics, and Astronomy, California Institute of Technology, Pasadena, California 91125, USA}%

\date{\today}

\begin{abstract}
Precision measurements of Schiff moments in heavy, deformed nuclei are sensitive probes of beyond Standard Model $T,P$-violation in the hadronic sector. While the most sensitive limits on Schiff moments to date are set with diamagnetic atoms, polar polyatomic molecules can offer higher sensitivities with unique experimental advantages.  In particular, symmetric top molecular ions possess $K$-doublets of opposite parity with especially small splittings, leading to full polarization at low fields, internal co-magnetometer states useful for rejection of systematic effects, and the ability to perform sensitive searches for $T,P$-violation using a small number of trapped ions containing heavy exotic nuclei.  We consider the symmetric top cation $^{225}$RaOCH$_3^+$ as a prototypical and candidate platform for performing sensitive nuclear Schiff measurements and characterize in detail its internal structure using relativistic \textit{ab initio} methods.  The combination of enhancements from a deformed nucleus, large polarizability, and unique molecular structure make this molecule a promising platform to search for fundamental symmetry violation even with a single trapped ion.
\end{abstract}

\maketitle

Searches for permanent electric dipole moments (EDMs) in atoms and molecules are powerful probes of time reversal and parity (\textit{T},\textit{P}-) violating physics posited by beyond Standard Model (BSM) theories \cite{safronova_search_2018,chupp_electric_2019}. Unsuppressed \textit{T},\textit{P}-violation, and by extension charge conjugation-parity (\textit{CP}-) violation, are needed to explain the observed lack of free antimatter in the universe \cite{dine_origin_2003}. In the Standard Model, however, $CP$-violation only weakly manifests in quark and neutrino mixing phases and is apparently absent for strong interactions (“the strong CP puzzle”) \cite{cheng_strong_1988}. The hadronic sector thus provides a natural venue for introducing many new \textit{CP}-violating BSM interactions to resolve this discrepancy \cite{yamanaka_probing_2017,engel_electric_2013,ginges_violations_2004}.

New \textit{T}, \textit{P}-violating nuclear effects, including nucleon-nucleon interactions mediated by QCD, are understood to induce a collective EDM in atomic nuclei of non-zero spin known as a Schiff moment \cite{schiff_measurability_1963,sushkov_possibility_1984, flambaum_nuclear_2002}. This effect, which scales with atomic mass $Z$, is particularly pronounced in heavy, octopole-deformed nuclei, such as $^{225}$Ra \cite{parker_first_2015,bishof_improved_2016}, where low-lying nuclear states couple strongly to the opposite-parity ground state \cite{gaffney_studies_2013}. The resulting Schiff moment, and corresponding sensitivity to BSM physics, is a factor of $\gtrsim 100$ larger \cite{auerbach_collective_1996,dobaczewski_nuclear_2005, dobaczewski_correlating_2018} when compared to heavy spherical nuclei, such as $^{129}$Xe \cite{sachdeva_new_2019,allmendinger_measurement_2019} and $^{199}$Hg \cite{graner_reduced_2016}, the latter of which is used in the current most sensitive Schiff moment experiment.

Heavy, octopole-deformed isotopes, however, are typically short-lived and difficult to produce in large quantities \cite{gaffney_studies_2013, butler_observation_2019, parker_first_2015}. Maximizing experimental sensitivity and coherence time is thus paramount to overcoming a limited count rate. One demonstrated method for increasing experimental sensitivity is to use a polar molecule, whose internal fields can be easily oriented to provide an enhancement of $\gtrsim 100$ over atoms in EDM measurements \cite{hudson_improved_2011,cairncross_precision_2017,andreev_improved_2018}. TlF, for instance, is sensitive to the Schiff moment of $^{205}$Tl nuclei \cite{cho_search_1991,hunter_prospects_2012}, and theoretical proposals have identified a wide variety of diatomic (ThO$^+$, ThF$^+$, AcF, AcO$^+$, AcN,  EuO$^+$, EuN, RaO, RaF) \cite{dzuba_electric_2002,flambaum_enhanced_2019,vskripnikov_actinide_2020,kudashov_calculation_2013, isaev_laser-coolable_2017,garcia_ruiz_spectroscopy_2020} and triatomic molecules (RaOH$^+$, TlOH, ThOH$^+$, TlCN) \cite{kozyryev_precision_2017, flambaum_enhanced_2019, kudrin_towards_2019} suitable for Schiff moment measurements. Combining enhancements due to nuclear deformation and the polarizability of molecules results in $\gtrsim 10^5$ sensitivity increase relative to atomic Schiff moment measurements with spherical nuclei. 

Molecular ions have proven to be a powerful platform for very sensitive measurements of symmetry violation  \cite{cairncross_precision_2017} due to long trapping and coherence times \cite{zhou_second-scale_2020}. This enables the ability to perform measurements with small quantities of the target molecule, for example those containing scarce or unstable nuclei.  However, many BSM-sensitive species, including radium, do not have the prerequisite electronic structure to make diatomic molecular ions with opposite-parity ($\Omega$) doublets, which are needed to fully realize the advantages of this approach.  Polyatomic molecules, by contrast, possess rovibrational parity doublets \cite{kozyryev_precision_2017}, and thus provide a generic approach to conducting molecular ion measurements with a broad range of useful, and possibly rare, species.  

In this manuscript, we consider a symmetric top molecule (STM), the radium monomethoxide cation (RaOCH$_3^+$), as a platform to combine nuclear and molecular enhancements with the advantages of a polyatomic structure and extended coherence time achievable with an ion trap. This molecule, which was recently produced and co-trapped \cite{fan_optical_2020} with laser-cooled Ra$^+$ \cite{fan_laser_2019}, has axial symmetry that gives rise to near-degenerate opposite parity $K$-doublets, thereby enabling full polarization in small fields and the co-magnetometer states necessary for sensitive measurements in an ion trap. The ground electronic state ($\tilde{X}^1\!A_1$) is diamagnetic, suppressing sensitivity to magnetic noise.
Due to this combination of enhancements and features, even a single trapped RaOCH$_3^+$ ion could be used to explore interesting parameter space for new physics.

There are several advantages to using a more complex STM ion, as opposed to a triatomic analog (e.g. RaOH$^+$) \cite{kozyryev_precision_2017,flambaum_enhanced_2019, maison_search_2020}. First, the increased rovibrational complexity of an STM, which makes laser cooling of neutral species more difficult (though indeed possible \cite{mitra_direct_2020, kozyryev_determination_2019}), does not pose challenges for the control of STM ions, as trapped ions do not require photon cycling to achieve high precision \cite{cairncross_precision_2017,zhou_second-scale_2020}. Furthermore, $K$-doublets, which arise from rotational degrees of freedom, can arise in any vibrational state, such as the ground state considered here.  They are therefore are low-lying ($\nu \sim 100$ GHz), have vastly longer radiative lifetimes than excited vibrational modes, and possess smaller splittings than the $\ell$-doublets of triatomics. 

Our theoretical analysis focuses on $^{225}$RaOCH$_3^+$, which contains the short-lived ($\tau_{1/2}\approx 15$~d) spin-$1/2$ radium isotope. We examine in detail the ground state hyperfine structure, as well as the various contributions to the degeneracy-breaking of the $K$-states. We furthermore identify states suitable for measurement of a Schiff moment, including co-magnetometer states, and examine the Stark and Zeeman effects in the molecule.

\begin{figure}
	\includegraphics[width=\columnwidth]{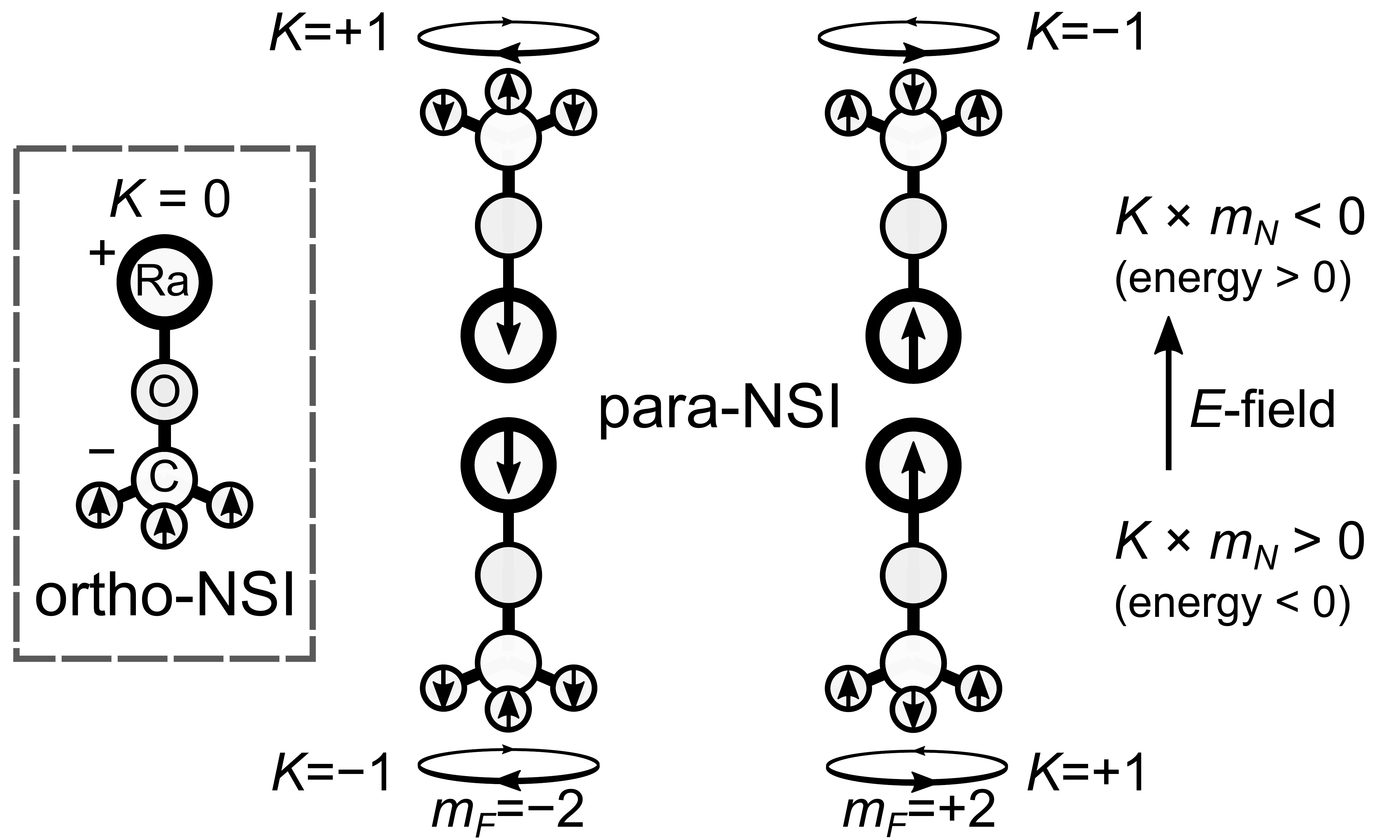}
	\caption{Labeled $^{225}$RaOCH$_3^+$ STMs in stretched states, grouped by energy and total angular momentum projection $m_F$.  The internuclear axis runs from the negatively charged end (methyl group) to the positively charged metal (radium). $K$ is the molecule-frame projection of angular momentum without spin $N$, while $m_N$ is its projection onto the lab frame. Nuclear spins for each spin-1/2 nucleus ($^1$H, $^{225}$Ra) are indicated. The $|K|=1$ states correspond to the mixed para nuclear spin isomer (NSI) of the hydrogens, while the $K=0$ state coincides with the stretched ortho-NSI.}
	\label{fig:blockdiagram}
\end{figure}

\textit{Internal Structure and $K$-doubling.} The internal structure of the electronic ground state ($\tilde{X}^1\!A_1$) is analyzed with explicit diagonalization of the effective molecular Hamiltonian
\begin{equation}
    H_{\text{total}} = H_{\text{rot}}+H_{\text{stark}}+H_{\text{zeeman}}+H_{\text{ss}}+H_{\text{nsr}}+H_{\text{sm}}.
\end{equation} 
We have included the rotational (rot), Stark, Zeeman, nuclear spin dipolar (ss), and nuclear spin-rotation terms (nsr), all of which are generic to STMs. The Schiff moment (sm) term arises from the $^{225}$Ra$(I=1/2)$ nucleus, and is \textit{T}, \textit{P}-violating. Similar to the closed-shell alkali monomethyls \cite{li_pure_1997}, electron spin terms are omitted. We obtain molecular parameters (see table \ref{tab:params}) using a variety of relativistic $\textit{ab initio}$ methods. Geometries are optimized at the level of CCSD(T) with an ANO-RCC-VQZ basis \cite{roos_main_2004, roos_relativistic_2004, widmark_density_1990, pritchard_new_2019,schuchardt_basis_2007, feller_role_1996} via CFOUR \cite{stanton_cfour_nodate, matthews_coupled-cluster_2020, harding_parallel_2008}, and scalar relativistic effects are modeled using the one-electron variant of the spin-free X2C Hamiltonian \cite{dyall_interfacing_1997, liu_exact_2009, cheng_analytic_2011}.  Nuclear spin-rotation and rotational Zeeman parameters are computed via a four-component relativistic linear response approach  \cite{aucar_theoretical_2012, aucar_theoretical_2013,aucar_theoretical_2014} in the DIRAC19 code  \cite{a_s_p_gomes_dirac_nodate} using the dyall.v4z basis \cite{dyall_relativistic_2009}, and electron correlation is treated at the DFT level with a B3LYP functional \cite{stephens_ab_1994}.  Additional details on the derivation of the Hamiltonian and \textit{ab initio} parameters can be found in the Supplemental Material.

The rotational structure of symmetric tops is parameterized by three quantum numbers: the electronic angular momentum apart from spin ($N$), its molecule-frame projection ($K$), and its lab-frame projection ($m_N$). For $|K|>0$, which corresponds to rotation about the symmetry axis, the cylindrical symmetry of the molecule gives rise to a pair of degenerate $+K$ and $-K$ states within each $|N, |K|\ket$ rotational manifold (see fig. \ref{fig:blockdiagram}). These degeneracies can be lifted by hyperfine and centrifugal terms that couple states of different $K$, leading to the formation of near-degenerate opposite parity $K$-doublets,  $|{\pm}\ket=(|N, +K\ket \pm |N, -K\ket)/\sqrt{2}$. 

We propose to use the $N=|K|=1$ manifold for the Schiff moment search. This state is $\sim160$ GHz above the absolute ground state, and accordingly has a much longer lifetime than any vibrationally excited mode (with frequencies $\gtrsim1$ THz) such as those in triatomics.  The spontaneous decay rate is further suppressed as the transition to the lower $K=0$ ground state is spin-forbidden, making the radiative lifetime much longer than 1 hour or any other relevant experimental timescale. 

\begin{figure}
	\includegraphics[width=0.9\columnwidth]{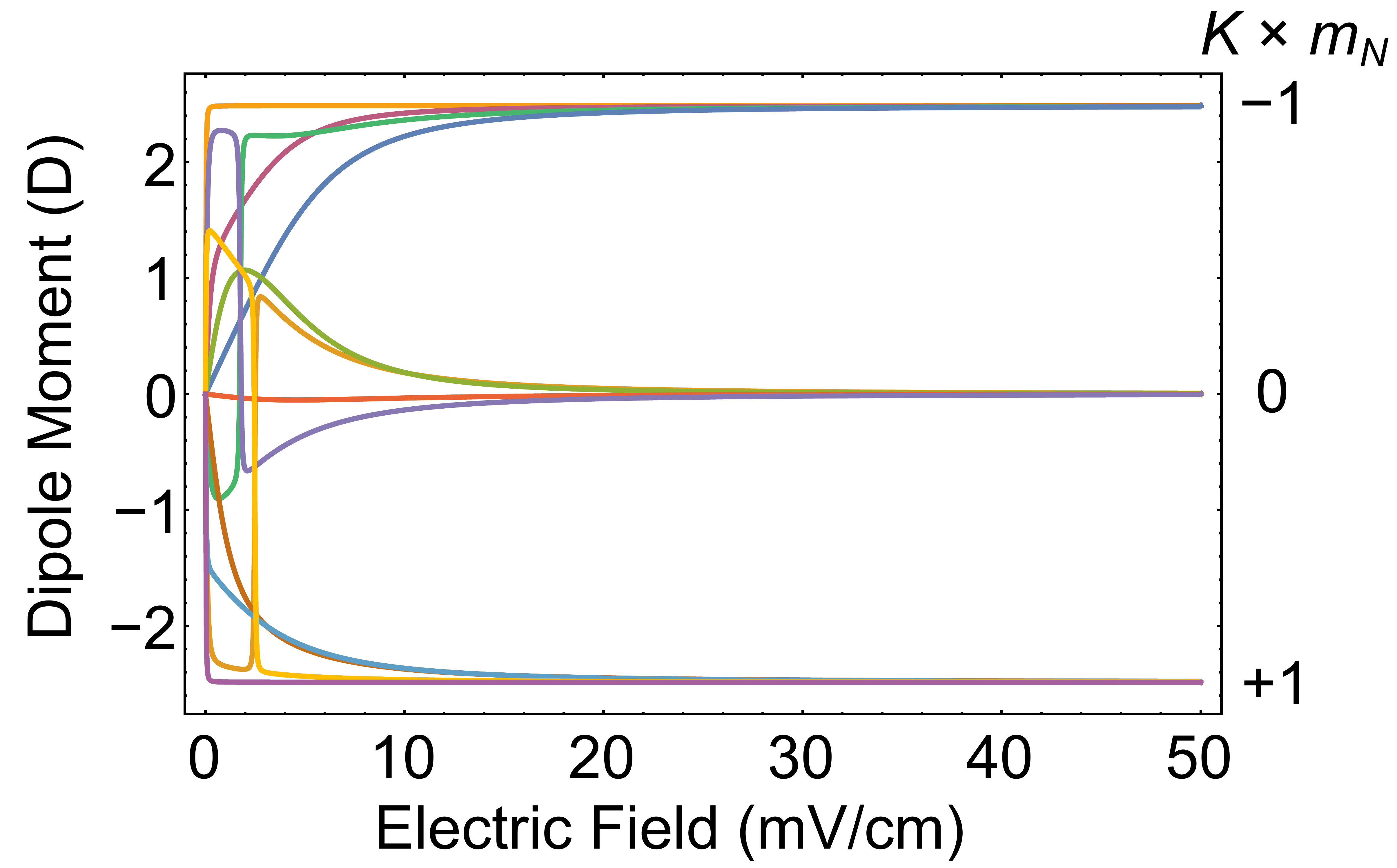}
	\caption{Lab frame dipole moment of hyperfine states of the $|N=1,|K|=1\ket$ manifold in the limit where the $K$ doublets are fully mixed yet rotational mixing is negligible. High, low, and no field seekers correspond to states with negative, positive, and zero dipole moment ($K\times m_N= +1$, $0$, and $-1$).	The jumps indicate avoided crossings. 
	}
	\label{fig:dipolefull}
\end{figure}

Similar to $\Omega$ and $\ell$-doublets, opposite parity $K$-doublets can be mixed in electric fields where the Stark energy exceeds the zero-field $K$-doublet splitting. In this regime, the molecule is polarized and its internal fields are oriented in the lab frame (see. fig \ref{fig:dipolefull}). This gives insensitivity to electric field fluctuations by largely saturating the Schiff moment sensitivity, as well as enabling co-magnetometer states.  In open-shell species, such as CaOCH$_3$, the splitting between the ground state $K$-doublets is $\sim 0.3$ MHz \cite{namiki_spectroscopic_1998}, dominated by an anisotropic hyperfine interaction between the proton spins and the metal-centered electron.  In contrast, the absence of unpaired electron spin in the $^{225}$RaOCH$_3^+$ ground state implies that the dominant hyperfine contributions to $K$-doubling are from nuclear spin interactions, which are suppressed generically by at least an order of magnitude due to the minute size of nuclear magnetic moments compared to electronic magnetic moments \cite{klemperer_can_1993,butcher_hyperfine_1993}. 

For the $N=|K|=1$ manifold,  we calculate that anisotropic nuclear spin-spin and spin-rotation contributions from the hydrogen nuclei generate sub-kHz $K$-doublings.  Combined with the calculated dipole moment of $\approx 5$~D, this results in an extremely low threshold for reaching the high-field limit and polarizing the molecule.  Indeed, we find that states in this manifold are $>90\%$ polarized in external electric fields of 50 mV cm$^{-1}$ and $>99.9\%$ polarized in fields of $250$~mV~cm$^{-1}$. This threshold is even lower for stretched states with maximal projection of total angular momentum ($m_F$), which reach full polarization ($>99.9\%$) in fields as low as $\lesssim$~mV~cm$^{-1}$ (see fig. \ref{fig:dipolefull}), small enough that the molecules could be polarized in $\sim$ mK deep optical traps \cite{schneider_optical_2010}.  Rotational mixing can be neglected at these small fields, which we assume is the case for the remainder of the manuscript.

\begin{figure}
	\includegraphics[width=\columnwidth]{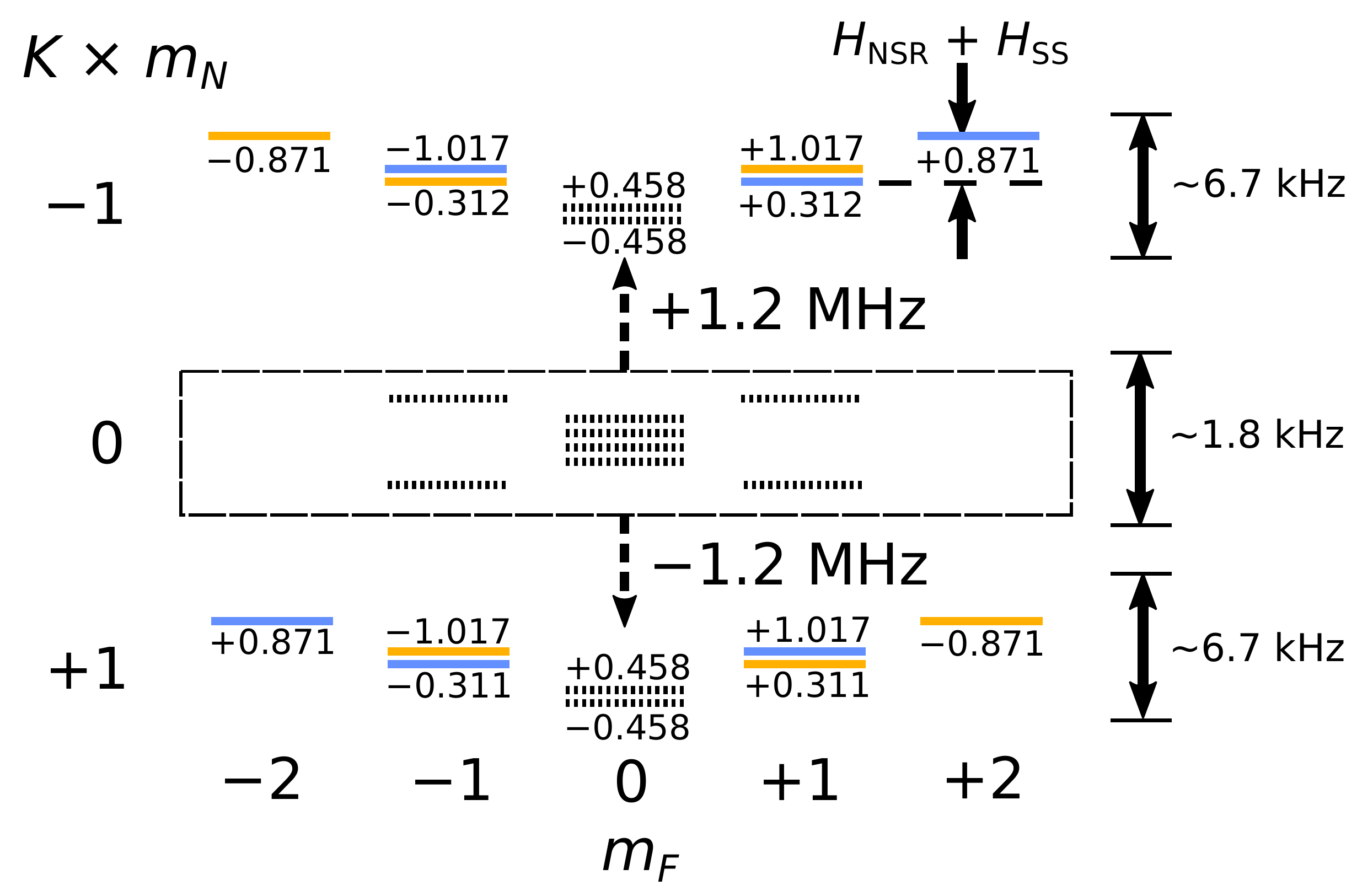}
	\caption{Level structure and Schiff moment sensitivities for the 24 hyperfine states of the $|N=1,|K|=1\ket$ manifold in the decoupled regime ($1$ V/cm), grouped by the projection of total angular momentum $m_F=m_N+m_{IH}+m_{IM}$ and their Stark manifold ($K\times m_N$). Gold states have +1/4 effective Schiff sensitivity, while blue states have $-1/4$ effective Schiff sensitivity. Dashed lines denote zero Schiff sensitivity. Labels above/below the states indicate the effective $g$-factor at zero magnetic field, in terms of nuclear magnetons ($\mu_N$).  Table \ref{tab:states_highfield} lists the admixtures for each state.}
	\label{fig:level_highfield_FmF}
\end{figure}

The absence of unpaired electron spin in the $\tilde{X}^1\!A_1$ ground state also has implications for the magnetic level structure, as only the nuclear and rotational moments contribute to the Zeeman energy of the ground state, and are of order $\sim\mu_N$, the nuclear magneton. This results in suppressed sensitivity to magnetic fields and effective $g$-factors that are a factor of $\sim10^3$ smaller than a Bohr magneton $\mu_B$ (see fig \ref{fig:level_highfield_FmF}). 

\textit{Hyperfine Structure.} The large number of internal degrees of freedom in $^{225}$RaOCH$_3^+$ necessitates a detailed treatment of the hyperfine structure to both identify resolvable states for the Schiff moment measurement as well as elucidate the sources of $K$-doubling. We use a fully decoupled basis, $|N, K, m_N\ket |\Gamma, I_H, m_{IH}\ket|I_{M}, m_{IM}\ket$, to describe the hyperfine structure of the $\tilde{X}\,^1\!A_1$ electronic state. The spin of the metal $^{225}$Ra nucleus is denoted $I_M$, and $m_{IM}$ is its corresponding lab frame projection. Similarly, the total nuclear spin of the three hydrogen atoms and its lab frame projection are denoted $I_H$ and $m_{IH}$, while $\Gamma$ denotes the symmetry character of the hydrogen spin wavefunction under $C_{3v}$ transformations. In concrete terms, these hydrogen spin wavefunctions correspond to either of two nuclear spin isomers (NSIs): the ``ortho" stretched states ($\Gamma=A, I_H=3/2$) and the ``para" mixed states ($\Gamma=E, I_H=1/2$). 

In each $|N, |K|\ket$ manifold, symmetry arguments restrict the allowed NSIs \cite{hougen_double_1980}. In particular, ortho states are only allowed with $K=3n$ rotational states (for integer $n$), while the para states are associated with $K\neq 3n$ states. All other combinations are forbidden by quantum statistics and $C_{3v}$ symmetries. (See Supplemental Material for details.) Accounting for these restrictions results in a total of 24 hyperfine states in the $N=1, |K|=1$ manifold, which are all resolvable in high fields.

Two types of hyperfine terms are present in the $\tilde{X}\,^1\!A_1$ state: dipolar couplings between the spins of different nuclei, and couplings between the nuclear spin and molecular rotation. The Hamiltonian for dipolar nuclear spin-spin interaction between two spins $\mathbf{I}_1$ and $\mathbf{I}_2$ can be expressed in terms of spherical tensors \cite{brown_rotational_2003}, 
\begin{equation}
    H_{\text{ss}}=-\sqrt{6}\frac{\mu_0\gamma_1\gamma_2\hbar^2}{4\pi}T^2(C_{\text{dip}})\cdot T^2(\mathbf{I}_1, \mathbf{I}_2).
\end{equation}
where $\mu_0$ is the vacuum permeability, $\gamma_1$, $\gamma_2$ are the gyromagnetic ratios, and $C_{\text{dip}}$ is a spin-spin coupling tensor. For the $N=|K|=1$ manifold, we only need to consider spin-spin interactions between the ortho-NSI hydrogens and the $^{225}$Ra atoms, as the inter-hydrogen matrix elements vanish between para-NSI states \cite{gunther-mohr_hyperfine_1954,wall_simulating_2013}. The dipolar spin coupling tensor $T^2(C_{\text{dip}})$, which can be directly evaluated as a sum of spherical harmonics, gives a diagonal shift of $\sim 40$ kHz for $\mathbf{I}_{H}\cdot \mathbf{I}_{R}$ interactions. Anistropic couplings of $\sim 400$ Hz mix states differing by $\Delta K=2$, which contributes to the $K$-doubling. 

Nuclear spin-rotation is the interaction between a nuclear magnetic moment associated with a spin $\mathbf{I}$ and the magnetic field created by the rotational angular momentum $\mathbf{N}$ \cite{hirota_high-resolution_1985},
\begin{equation}
    H_{\text{nsr}}=\frac{1}{2}\sum^2_{k=0}\big[T^k(C_{\text{nsr}})\cdot T^k(\mathbf{N}, \mathbf{I})+T^k(\mathbf{N}, \mathbf{I})\cdot T^k(C_{\text{nsr}}) \big],
\end{equation}
where $C_{\text{nsr}}$ is a spin-rotation coupling tensor. Both $^{225}$Ra and the hydrogen nuclei in the ortho-NSI contribute to the nuclear spin-rotation interaction.  The former produces a diagonal shift of $\sim 4$ kHz for $\mathbf{N}\cdot \mathbf{I}_R$ interactions, while the latter produces both a diagonal shift of $\sim 15$ kHz for $\mathbf{N}\cdot \mathbf{I}_H$ interactions and $\sim 300$ Hz off-diagonal couplings between states differing by $\Delta K=2$, which contributes to the $K$-doubling. 

\begin{table}[b]
\caption{\label{tab:params}%
Molecular parameters for $^{225}$RaOCH$_3^+$ ground state ($\tilde{X}^1\!A_1$). See Supplemental Material for details.
}
\begin{ruledtabular}
\begin{tabular}{llll}
\multicolumn{2}{l}{\textrm{Hyperfine}}&
\multicolumn{1}{c}{$T_{zz}$}&
$|T_{xx}-T_{yy}|$\\
\colrule
\multicolumn{2}{l}{$T(C_{\text{nsr}}(^{225}\textrm{Ra}))$} & $\phantom{-}3.67$ kHz & -- \\
\multicolumn{2}{l}{$T(C_{\text{nsr}}(^1\textrm{H}))$} & $\phantom{-}15.3$ kHz& $0.301$ kHz \\
\multicolumn{2}{l}{$\alpha_{\text{dip}}\cdot T(C_{\text{dip}}(^{225}\textrm{Ra}-\textrm{H}))$\footnote{The scaling constant for the dipolar spin-spin interaction is defined as $\alpha_{\text{dip}}=-\sqrt{6}\mu_0\gamma_{\text{H}}\gamma_{\text{Ra}}\hbar^2/4\pi$}} & $-38.0$ kHz & $0.391$ kHz \\
\hline
\textrm{Geometry}& & \textrm{Stark and Zeeman} &\\
\hline
r(Ra-O) & {2.1949 \AA} & $d_0$ & \multicolumn{1}{r}{4.969 D}\\
r(O-C) & {1.4076 \AA} & $g_N(^{225}\textrm{Ra})$\textsuperscript{\cite{arnold_direct_1987}} & {$-0.7338\mu_N$}\\
r(C-H) & {1.0864 \AA} & $g_N(\textrm{H})$\textsuperscript{\cite{singh_nuclear_2005}} & \multicolumn{1}{r}{$2.7928\mu_N$}\\
$\angle$(O-C-H) & {110.73\degree} & $g_{R(\parallel)}$ & \multicolumn{1}{r}{  $6.65\mu_N$}\\
$\angle$(H-C-H) & {108.18\degree} & $g_{R(\perp)}$ & \multicolumn{1}{r}{$0.619\mu_N$} \\
$A$ & {5.4010 cm$^{-1}$}& &\\
$B$ & {0.0673 cm$^{-1}$}& &\\
\end{tabular}

\end{ruledtabular}
\end{table}

\textit{Measurement.} Since the energy shift from the Schiff moment is proportional to the projection of the radium spin onto the molecular axis ($\mathbf{I}_M\cdot \mathbf{\hat{n}}$), different hyperfine states have different Schiff moment sensitivities. In the fully decoupled limit ($\geq 1$ V/cm), the value of $\bra \mathbf{I}_M\cdot \mathbf{\hat{n}} \ket$ is $K \cdot m_N\cdot m_{IM}/2$, where the prefactor arises because we do not mix rotational states beyond $N=1$ and thus the maximum projection of the internuclear axis on the lab frame is $1/2$. (see fig. \ref{fig:level_highfield_FmF} and table \ref{tab:states_highfield}). There are therefore three distinct classes of states, with positive, zero, and negative Schiff energy shifts. Even in the intermediate regime when a molecule is polarized, but the spins are not decoupled ($\sim 50$ mV cm$^{-1}$), we still find stretched states that maximally project the radium spin onto the molecular axis.

The measurement manifold is naturally populated at cold temperatures, with $\sim 1\%$ of molecules occupying the $N=|K|=1$ manifold at 4 K. This yield could be increased via state-controlled reactions of Ra$^+$ and methanol \cite{hu_product-state_2020}. State preparation is possible through state-selective dissociation \cite{zhou_second-scale_2020} or non-destructive quantum logic \cite{patterson_method_2018, chou_preparation_2017} via co-trapped Ra$^+ $\cite{fan_laser_2019}.

Two schemes can be used for performing the EDM measurement. In the spin interferometery (SI) method \cite{cairncross_precision_2017}, the molecule precesses between states of different Schiff sensitivity. After a time $\tau$, the phase $\phi=(\omega_B+\omega_{TP})\tau$ is extracted via projective measurements, where $\omega_B\propto g_{\text{eff}}\,\mu_N|B_0|/\hbar$ is the Larmor precession frequency and $\omega_{TP}\propto \Delta H_\text{sm}/\hbar$ is the contribution from the differential Schiff moment between the two states. The $\Delta H_\text{sm}$ contribution to the phase can be distinguished from the Larmor precession by repeating the measurement in different magnetic fields and with different relative orientations of the radium nuclear spin and internuclear axis, which is enabled in this case by the $K$-doublets. A proposed alternative approach, known as the clock-transition (CT) method  \cite{verma_permanent_2020}, uses time-dependent electromagnetic fields to drive transitions between different hyperfine ``clock" states. The $T,P$-violating interaction is then extracted from phase-dependent shifts to the measured Rabi oscillations. This technique, which is readily adapted to an ion trap, benefits from a simpler state preparation scheme and better robustness to electromagnetic noise.

In the polarized limit, the rotational manifold contains many hyperfine states for driving the SI or CT measurement scheme. Each state has different effective $g$-factors and Schiff sensitivity (see Fig. \ref{fig:level_highfield_FmF}). Performing an EDM measurement with pairs of states which have unique differential magnetic sensitivities enables one to adjust the Larmor precession without changing the applied $B$-field.  In addition, there are  multiple pairs of near-degenerate states of opposite $m_F$ with the same Schiff moment sensitivity, but different magnetic moments. These add to the set of valuable systematic checks.

\textit{BSM Sensitivity.}  Calculations of the Schiff moment of $^{225}$Ra nuclei have been performed in the framework of $T,P$-violating pion exchange between nucleons \cite{dobaczewski_nuclear_2005}, yielding parameterizations of $\mathbf{S}(^{225}\text{Ra})$ in terms of the QCD $\bar{\theta}$ angle given by $|\mathbf{S}(^{225}\text{Ra})|=$ $1.0 \bar{\theta} \text{ \textit{e} fm}^3$ \cite{flambaum_enhanced_2020}. The electrostatic interactions generated by the Schiff moment leads to an effective $T,P$-violating shift $H_{\text{sm}}=W_s\, (\mathbf{I}\cdot \mathbf{\hat{n}})\,|\mathbf{S}|/|\mathbf{I}|$. The species-dependent coupling constant $W_s$, which is an electron-nuclear contact term, has been calculated to be $45,192$ atomic units in RaO \cite{kudashov_calculation_2013} and is estimated to be slightly smaller ($\sim 30,000$ a.u.) for RaOH$^+$ (a.u.  $\equiv e/4\pi\epsilon_0 a_0^4$), where the larger ligand is assumed to reduce both electron density around Ra and the magnitude of $W_s$ \cite{flambaum_enhanced_2019}.

To illustrate the power of a Schiff moment measurement on $^{225}$RaOCH$_3^+$, we can combine the QCD parameterization of $\mathbf{S}(^{225}\text{Ra})$ with the estimate $W_s(^{225}$RaOCH$_3^+)\approx 30,000$ a.u. to calculate the averaging time needed to reach a new model-dependent limit on QCD $\bar{\theta}$. 
We assume a single trapped $^{225}$RaOCH$_3^+$ ion with 5 s coherence time limited by black-body pumping at 300 K, which provides a frequency sensitivity of $\delta\omega=7.5$ mrad s$^{-1}/\sqrt{\text{hour}}$. Spin-precession measurements with this setup would reach a statistical sensitivity $\bar\theta<10^{-10}$ with two weeks of data taking.  Trapping multiple ions and improving the coherence time through cryogenic cooling would result in even higher sensitivity.

\textit{Conclusion and outlook.} We have considered trapped $^{225}$RaOCH$_3^+$ as a sensitive platform to search for a nuclear Schiff moment in the octopole-deformed Ra nucleus. While our theoretical calculations do not replace the need for detailed spectroscopic studies on this particular species, they do illustrate advantageous structures that are quite general for both symmetric and asymmetric top molecules \cite{augenbraun_molecular_2020,yu_scalable_2019,wall_simulating_2013}.  This approach can therefore be used to search for a variety of fundamental symmetry violations in many different species, including those with exotic nuclei such as Pa \cite{flambaum_electric_2008} and U \cite{ wormit_strong_2014}, and many ligands including chiral species.

Over the years, a wide variety of metal-monohydroxide (MOH$^+$), metal-monomethoxide (MOCH$_3^+$), as well as dual-metal hypermetallic (MOM$'^+$) ions have been created \cite{lu_reactions_1998,lu_reactions_1998-1} (and, in some cases, trapped \cite{puri_synthesis_2017, yang_optical_2018}), including species with heavy nuclei such as Ba \cite{puri_synthesis_2017} and Lu \cite{lu_reactions_1998-1} in addition to Ra \cite{fan_optical_2020}. The rich internal complexity of these molecules makes them attractive for a broad range of studies not limited to Schiff moment measurements \cite{safronova_search_2018}.  Much of the discussion in this manuscript, for instance, is directly applicable to searches for the electron EDM or nuclear magnetic quadrupole moments \cite{kozyryev_precision_2017,maison_search_2020}.  The availability of opposite parity states with diverse, tunable splittings is particularly useful for precision measurement of electroweak physics, such as nuclear spin-dependent parity violation \cite{norrgard_nuclear-spin_2019} and oscillating symmetry violations from interactions with axion-like fields \cite{graham_axion_2011,stadnik_axion-induced_2014,flambaum_oscillating_2019}, and the sources that generate the splittings can be sensitive to variations of fundamental constants \cite{jansen_perspective_2014,kozyryev_enhanced_2018}.

\textit{Acknowledgments.}  We are grateful for extensive assistance from Anastasia Borschevsky and Y.A. Chamorro Mena with the \emph{ab initio} calculations, and to Ben Augenbraun, Mingyu Fan, Alex Frenett, Arian Jadbabaie, Andrew Jayich, Ivan Kozyryev, Zack Lasner, and Tim Steimle for helpful discussions and feedback. This research was supported by a NIST Precision Measurement Grant (60NANB18D253), the Gordon and Betty Moore Foundation (7947), and the Alfred P. Sloan Foundation (G-2019-12502). Computations in this manuscript were performed on the Caltech High Performance Cluster.

\bibliography{ref}

\newpage
\begin{center}
\textbf{\large Supplemental Material}
\end{center}

\setcounter{section}{0}
\setcounter{equation}{0}
\setcounter{figure}{0}
\setcounter{table}{0}
\makeatletter
\renewcommand{\theequation}{S\arabic{equation}}
\renewcommand{\thefigure}{S\arabic{figure}}
\renewcommand{\thetable}{S\arabic{table}}
\renewcommand{\thesection}{S\arabic{section}}

\section{Molecular Structure}

As discussed in the main text, we numerically diagonalize the effective molecular Hamiltonian for the ground electronic state ($\tilde{X}^1\!A_1$) of $^{225}$RaOCH$_3^+$
\begin{equation}
    H_{\text{total}} = H_{\text{rot}}+H_{\text{stark}}+H_{\text{zeeman}}+H_{\text{ss}}+H_{\text{nsr}}+H_{\text{sm}}
\end{equation} 
where we have included the rotational (rot), Stark, Zeeman, nuclear spin dipolar (ss), nuclear spin-rotation terms (nsr), and Schiff moment (sm) terms. No electron spin terms are included, as the molecule has a closed shell. For generality, however, the  matrix elements are written in the fully decoupled basis including the electron spin $S$: $|N, K , S, J, m_J\ket|I_M, m_{IM}\ket|\Gamma, I_H, m_{IH}\ket$.

\subsection{Rovibrational Structure}
RaOCH$_3^+$ is a prolate symmetric top with point group $C_{3v}$, corresponding to its three-fold cylindrical symmetry about the principal molecular axis $(Z)$. The Hamiltonian that corresponds to the rotational energy for a prolate top is
\begin{align}
    H_{\text{rot}}=B\mathbf{N}^2 + (A-B)\mathbf{N}_Z^2,
\end{align}
which has eigenenergies $BN(N + 1) + (A-B)K^2$. $A$ and $B$ are rotational constants, while $\bra \mathbf{N}\ket=N$ and $\bra \mathbf{N}_Z\ket=K$ are the canonical rotational quantum numbers.  $A$ corresponds to the rotation of the molecule about the symmetry axis, while $B$ corresponds to end-over-end rotation. For $|K|\neq 0$, each $|N,|K|\ket$-level in the rotational Hamiltonian has $2\times (2N + 1)$ degeneracies, which are given by the $2N+1$ lab-frame projections $m_N$ and the two projections of $|K|$ onto the molecular axis.

The \textit{ab initio} geometries, with corresponding rotational constants, for RaOCH$_3^+$ (see table \ref{tab:params} in main text) were optimized using the coupled cluster method with single, doubles, and perturbative triples [CCSD(T)] \cite{raghavachari_fifth-order_1989, bartlett_non-iterative_1990} in CFOUR \cite{stanton_cfour_nodate, matthews_coupled-cluster_2020, harding_parallel_2008}. Relativistically contracted atomic natural orbital basis sets of polarized double, triple, and quadruple-zeta quality [ANO-RCC-V$n$ZP ($n$=D,T,Q)] are used \cite{widmark_density_1990, roos_main_2004, roos_relativistic_2004, pritchard_new_2019, feller_role_1996, schuchardt_basis_2007}. Scalar relativistic effects are included via the one-electron variant of the spin-free exact two component theory (SFX2C-1e) \cite{dyall_interfacing_1997, liu_exact_2009, cheng_analytic_2011}. 

There are eight vibrational modes for symmetric top molecules of the MOCH$_3$ form, four symmetric modes of $a_1$ character and four asymmetric modes of $e$ character. The vibrational energies and intensities of each mode are calculated via analytic B3LYP Hessians \cite{bykov_efficient_2015,stephens_ab_1994} with ORCA \cite{neese_orca_2012}. For this calculation, we employ correlation-consistent basis sets at the quadruple-zeta level \cite{dunning_gaussian_1989}, with core-valence \cite{hill_gaussian_2017} and pseudopotential sets for the radium atom. The 78 core electrons of radium are modeled using the SK-MCDHF-RSC effective core potential (ECP) \cite{lim_relativistic_2006}.  Energies and symmetry characters are listed in Table \ref{tab:vib}. While the proposed Schiff moment search would take place in the ground vibrational state, excited ro-vibrational states could be a valuable resource for state preparation or readout \cite{kondov_molecular_2019,manai_rovibrational_2012,shimasaki_production_2015,staanum_rotational_2010,khanyile_observation_2015}.  The energy difference between the nominally degenerate states of $e$ character indicate that the accuracy of the energies is on the few percent level.

\begin{table}
    \centering
    \begin{ruledtabular}
    \begin{tabular}{lll}
        Vibration & Energy [cm$^{-1}$] & Character\\
        \hline 
        Ra-O-C bend & $164.96/168.68$ & $e$\\
        Ra-O stretch & $390.78$ & $a_1$\\
        CH$_3$ symmetric bend & $1086.84$ & $a_1$\\
        CH$_3$ rock & $1170.41/1173.36$ & $e$\\
        C-O stretch & $1481.50$ & $a_1$\\
        CH$_3$ asymmetric bend & $1497.61/1498.56$ & $e$\\
        CH$_3$ symmetric stretch & $2993.19$ & $a_1$\\
        CH$_3$ asymmetric stretch & $3048.22/3059.15$ & $e$\\
    \end{tabular}
    \end{ruledtabular}
    \caption{Vibrational energies, computed from B3LYP Hessians \cite{bykov_efficient_2015} at DFT optimized geometry, where each mode is classified with respect to its transformations under $C_{3v}$ symmetry. In total, there are four symmetric $a_1$ states and four doubly-degenerate $e$ states. The pair of frequencies for the degenerate vibrations is due to slight symmetry-breaking in the computed geometry.}\label{tab:vib}
\end{table}

The vibrational states are also relevant as they will likely present a limitation on the coherence time \cite{leanhardt_high-resolution_2011}.  Black-body excitation of the Ra-O stretch mode (with the transition dipole moment calculated to be $\mu\sim 0.26$ D) is estimated to occur with a $\sim$5 second time scale in a 300 K environment, though that can be reduced to 20 minutes in a 77 K environment.  The radiative lifetime of the $N=|K|=1$ states are much longer than one day due to their small energy spacing (a state with one atomic unit of transition dipole moment at this frequency would last around one hour) and the fact that they are spin-forbidden to decay to the ground rotational state, so black-body effects are likely to dominate over radiative decay at room temperature.

\subsection{Molecular Symmetries}
\subsubsection{Classification of Rovibronic and Nuclear Spin Wavefunctions}
The $C_{3v}$ point group (which is isomorphic to $S_6$) has three irreducible representations: $A_1$, $A_2$, and $E$. $A_1$ denotes the fully symmetric representation, and $A_2$ is the anti-symmetric representation. $E$ denotes the degenerate representation, which splits into positive ($E_{+}$) and negative ($E_{-}$) components under the action of the cyclic group $C_3\subset C_{3v}$. The character tables for $C_{3v}$ and $C_{3}$ are written in Table \ref{tab:chars} for reference. 
\begin{table}[h]
    \centering
    \begin{tabular}{c|c|c|c|}
              & $\hat{E}$ & $2\hat{C}_3(z)$ & $3\sigma_v$ \\
        \hline 
        $A_1$ & $+1$ & $+1$ & $+1$\\
        \hline
        $A_2$ & $+1$ & $+1$ & $-1$ \\
        \hline
        $E$   & $+2$ & $-1$ & $0$ \\
        \hline
    \end{tabular}\qquad
    \begin{tabular}{c|c|c|c|}
              & $\hat{E}$ & $\hat{C}_3$ & $\hat{C}_3^2$ \\
        \hline 
        $A$ & $+1$ & $+1$ & $+1$\\
        \hline
        $E_{\pm}$ & $+1$  & $e^{\mp 2\pi i/3}$ & $e^{\pm 2\pi i/3}$ \\
        \hline
    \end{tabular}
    \caption{Character Tables for $C_{3v}$ (left) and $C_3$ (right)}\label{tab:chars}
\end{table}

We start by classifying the rovibronic states. In the electronic ground state and a vibrationally relaxed manifold ($|\Lambda=0, \ell=0\ket$), the symmetry classification of the $|N,|K|\ket$ rovibronic states is dependent only on $K$. The $|N,K=0\ket$ state transforms as $A_1$ or $A_2$ (depending on $N$), while $|N,|K|=3n\ket$ states, where $n$ is an integer and $K\neq 0$, transform as $A_1\oplus A_2$. The remaining $|N,|K|\neq 3n\ket$ states transform as the doubly degenerate $E_{\pm}$ character, where the positive and negative components correspond to the positive and negative projections of $|K|$. (A generalized classification for non-zero orbital and vibrational angular momentum ($|\Lambda\neq 0, l\neq 0\ket$) can be found in \cite{hougen_double_1980}.) The dependence of the molecular symmetries on $K$ arises from the fact that the hydrogen atoms in the methyl group are indistinguishable and must obey the Pauli principle, as discussed in detail later.

All states where $|K|>0$ are doubly degenerate and thus correspond to either the mixed $E_{\pm}$ or $A_1\oplus A_2$ symmetry characters. The splitting of $K$-doublets (and thus the breaking of $C_{3v}$ symmetry) is naturally associated with the splitting of $E_{\pm}$ or $A_1\oplus A_2$ into distinct representations.

We discuss the specific mechanisms lifting the degeneracy below, but there is a simple and intuitive picture how this arises due to hyperfine couplings \cite{gunther-mohr_hyperfine_1954}. The $K$-doubled states $|\pm\ket$ have a different distribution of proton spin about the azimuthal angle relative to the symmetry axis and therefore the anisotropic nature of the dipolar or nuclear spin-rotation interaction splits the two states. For $|K|=1$ states, the hyperfine anisotropies directly couple states differing by $\Delta K=2$ to produce a first-order $K$-splitting, but have progressively suppressed effects on the splitting for higher $|K|>1$ \cite{gunther-mohr_hyperfine_1954,wofsy_hyperfine_1970}.

For states where $|K|=3n$, the leading order source of $K$-doubling arises from a sextic centrifugal distortion term, which couples $\Delta K=6$. This term appears as a correction to the rotational Hamiltonian $H_\text{sextic}=q_3(J_+^6+J_-^6)/2$, where $q_3$ is the distortion constant \cite{aliev_calculated_1976}. In the CH$_3$ radical, the $q_3$ distortion constant has been measured to be $ 370$ Hz \cite{kawaguchi_high-resolution_2001}. At higher multiples of three, the $K$-splitting generated by $q_3$ is again suppressed. Using higher $K$ states is therefore a general way to obtain even smaller $K$-doublets.

The composite nuclear spin states $|\Gamma, I_H, m_{IH}\ket$ of the three hydrogen atoms, where $\Gamma$ denotes the symmetry, are also classified into four states that transform as $A_1$, and four states that transform as $E_{\pm}$:
\begin{widetext}
\begin{align}
    \Line A, \frac{3}{2}, \pm\frac{3}{2}\Ket&=\Line \pm\frac{1}{2}, \pm\frac{1}{2}, \pm\frac{1}{2}\Ket\\
    \Line A, \frac{3}{2}, \pm\frac{1}{2}\Ket&=\frac{1}{\sqrt{3}}\bigg(\Line \pm\frac{1}{2}, \pm\frac{1}{2}, \mp\frac{1}{2}\Ket+\Line \pm\frac{1}{2}, \mp\frac{1}{2}, \pm\frac{1}{2}\Ket+\Line \mp\frac{1}{2}, \pm\frac{1}{2}, \pm\frac{1}{2}\Ket\bigg)\\
    \Line E_+, \frac{1}{2}, \pm\frac{1}{2}\Ket&=\frac{1}{\sqrt{3}}\bigg(\Line \mp\frac{1}{2}, \pm\frac{1}{2}, \pm\frac{1}{2}\Ket+e^{+ 2\pi i /3}\Line \pm\frac{1}{2}, \mp\frac{1}{2}, \pm\frac{1}{2}\Ket+e^{- 2\pi i /3} \Line \pm\frac{1}{2}, \pm\frac{1}{2}, \mp\frac{1}{2}\Ket\bigg)\\
    \Line E_-, \frac{1}{2}, \pm\frac{1}{2}\Ket&=\frac{1}{\sqrt{3}}\bigg(\Line \mp\frac{1}{2}, \pm\frac{1}{2}, \pm\frac{1}{2}\Ket+e^{- 2\pi i /3}\Line \pm\frac{1}{2}, \mp\frac{1}{2}, \pm\frac{1}{2}\Ket+e^{+2\pi i /3}\Line \pm\frac{1}{2}, \pm\frac{1}{2}, \mp\frac{1}{2}\Ket\bigg)
\end{align}
\end{widetext}
$I_H$ refers to the total composite spin of the three protons ($\mathbf{I}_H=\mathbf{I}_{H1}+\mathbf{I}_{H2}+\mathbf{I}_{H3}$) and $m_{IH}$ is the corresponding lab frame projection. $\Gamma$ denotes the symmetry of the composite spin state.

To obey Fermi-Dirac statistics, the total nuclear-rovibronic wavefunction must transform as either $A_1$ (symmetric under inversion) or $A_2$ (antisymmetric under inversion) \cite{hougen_double_1980}. Therefore, the $K=1$ and $K=2$ (in the ground state) cases correspond to combined nuclear ($^n\Gamma$) and rovibronic ($^{evsr}\Gamma$) wavefunctions that have the symmetries
\begin{align}
    |^{evsr} E_+\ket|^n E_-\ket\hspace{0.25 cm}\text{or}\hspace{0.25 cm}
    |^{evsr} E_-\ket|^n E_+\ket,\label{eq:totalsymE}
\end{align}
whereas the fully symmetric $K=0$ or $K=3$ rotational states correspond to two possibilities
\begin{align}
    |^{evsr} A_1\ket|^n A\ket\hspace{0.25 cm}\text{or}\hspace{0.25 cm}
    |^{evsr} A_2\ket|^n A\ket.\label{eq:totalsymA}
\end{align}
From eq. (\ref{eq:totalsymE}), we can observe that any hyperfine interaction which couples $|^nE_{+}\ket$ and $|^nE_{-}\ket$ states also couples the $|^{evsr} E_+\ket$ and $|^{evsr} E_-\ket$ states, and therefore splits any residual $K$-degeneracy into the doublets,
\begin{align}
    \frac{1}{\sqrt{2}}(|^{evsr} E_+\ket|^n E_-\ket\pm |^{evsr} E_-\ket|^n E_+\ket).
\end{align}
This is the case for the $K$-doubling that originates from the nuclear dipolar spin couplings of RaOCH$_3^+$, for instance. 

The symmetry assignments in eqs. (\ref{eq:totalsymE}) and (\ref{eq:totalsymA}) also lead to additional selection rules for electric dipole transitions. For instance, let us consider transitions between different $|N,K\ket$ manifolds in the electronic-vibrational ground state. Starting from the $K=1$ manifold, the electric dipole operator only couples to $\Delta K = 0, \pm 1 $ states. $|N, K=2\ket$ has the same nuclear symmetry $\Gamma$ as $| N, K=1\ket$ and the transition is allowed. $|N,K=0\ket$ does not, and the transition is symmetry-forbidden (as well as spin-forbidden), which is why the radiative lifetime of $|N,K=1\ket$ is so long. 

\begin{figure*}
	\includegraphics[width=\columnwidth]{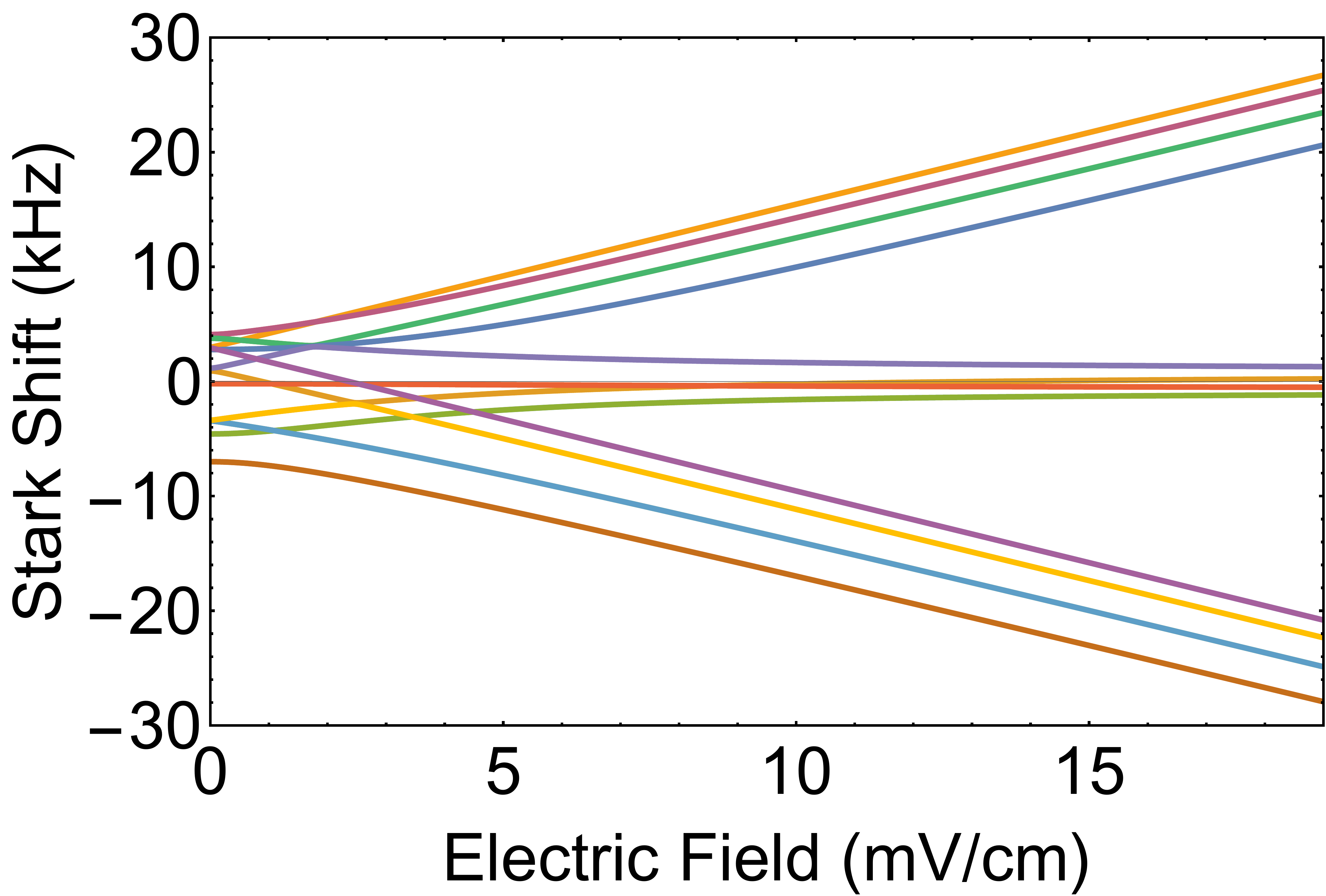}
	\includegraphics[width=\columnwidth]{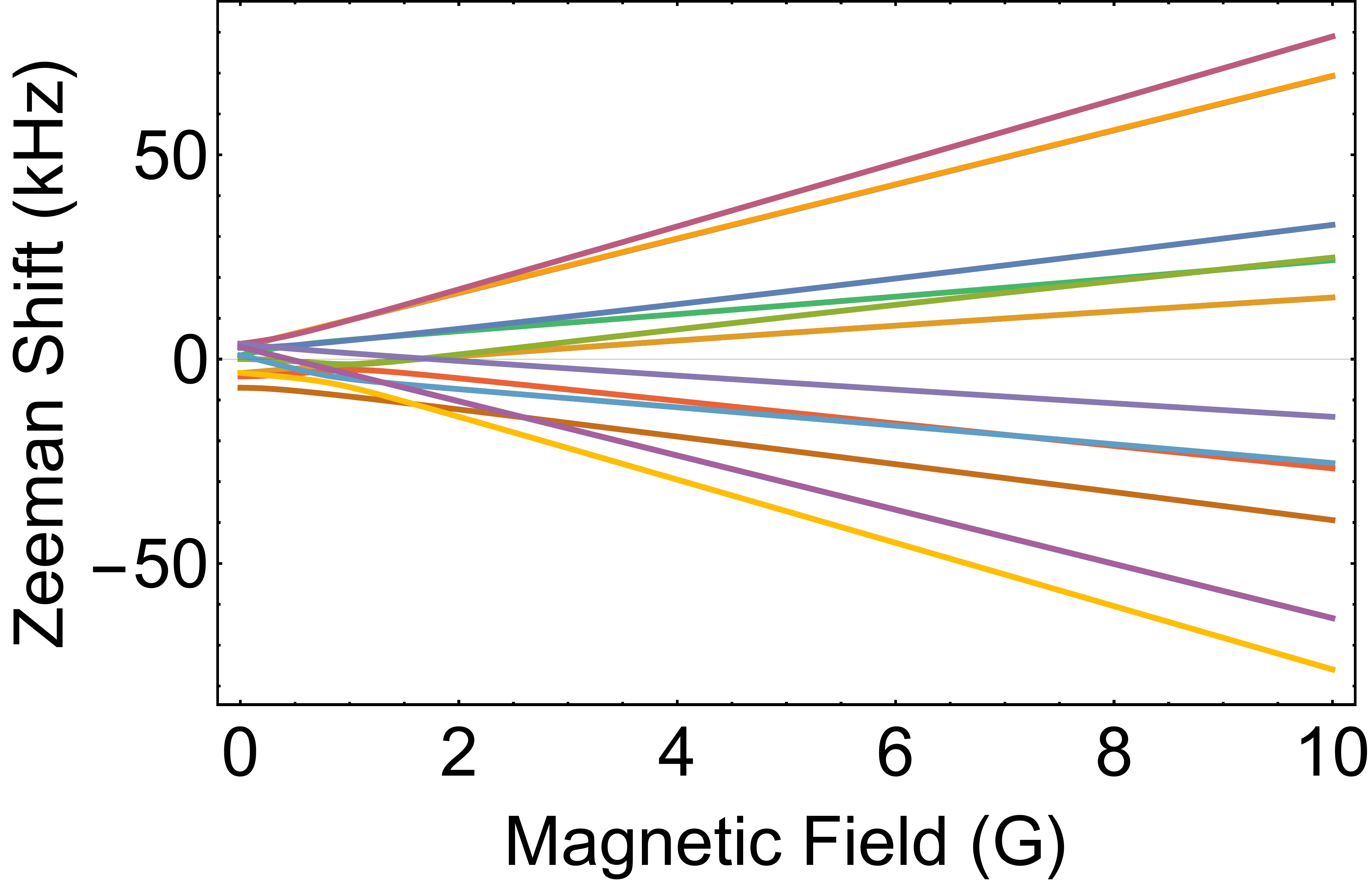}
	\caption{(left) Stark plot of hyperfine states in the $|N=1,|K|=1\ket$ manifold, up to 15 mV cm$^{-1}$ electric field. (right) Zeeman plot of hyperfine states in the $|N=1,|K|=1\ket$ manifold, up to 10 G magnetic field.}
	\label{fig:starkzeeman}
\end{figure*}

\subsection{Hyperfine Structure}
\subsubsection{Parameterization of the Tensor Operators}
In order to evaluate hyperfine structure of the hydrogen spins, it is necessary to sum over the interactions involving each individual individual hydrogen nucleus. A spin-spin interaction between the hydrogen spins $I_i$ and an arbitrary spin $\mathcal{S}$, for instance, has the form
\begin{align}
    \sum_{i=1}^3 \mathcal{S}\cdot T_i\cdot I_i,\label{unparametrizedint}
\end{align}
where $T_i$ is the interaction tensor between the $i$th hydrogen and $\mathcal{S}$. In Cartesian form, $T_1$ can be written as
\begin{align}
    T_1=\begin{pmatrix}{}
    T_{xx} & 0 & T_{xz}\\
    0 & T_{yy} & 0\\
    T_{zx} & 0 & T_{zz}\\
    \end{pmatrix} \label{eq:t1},
\end{align}
while $T_2$ and $T_3$ can be obtained with the appropriate rotations about the $z$ axis \cite{endo_microwave_1982}.

The form of eq. (\ref{unparametrizedint}), however, is unwieldy for evaluating matrix elements over the symmetrized nuclear spin states derived in the previous section. A more intuitive form can be obtained by parameterizing the $T$ and $I$ operators, as first formulated by Hougen in cartesian form \cite{hougen_double_1980} and Endo et al. in spherical tensor form \cite{endo_microwave_1982, endo_microwave_2003}.
\begin{align}
    T_0&= (T_1+T_2+T_3)/3,\\
    T_\pm&= (T_1+e^{\pm 2\pi i/3}T_2+e^{\mp 2\pi i/3}T_3)/3,\\
    I_0&= I_1+I_2+I_3,\\
    I_\pm&= I_1+e^{\pm 2\pi i/3}I_2+e^{\mp 2\pi i/3}I_3.
\end{align}
This allows us to rewrite eq. (\ref{unparametrizedint}) as
\begin{align}
    \sum_{\alpha=0, \pm} S\cdot T_\alpha\cdot I_{-\alpha}.
\end{align}
Direct evaluation via the Wigner-Eckart Theorem yields the following reduced matrix elements for the parameterized nuclear spin operators over the symmetrized nuclear spin basis:
\begin{align}
    &\bra \Gamma, I_0|T^1(\mathbf{I}_0)|\Gamma, I_0\ket=\sqrt{I_0(I_0+1)(2I_0+1)},\label{eq:reparam1}\\
    &\bra E_{\pm}, I_0|T^1(\mathbf{I}_\pm)|E_{\mp}, I_0\ket=-2\sqrt{I_0(I_0+1)(2I_0+1)},\label{eq:reparam2}\\
    &\bra E_{\pm}, I_0=1/2|T^1(\mathbf{I}_\pm)|A_1, I_0=3/2\ket=\sqrt{6}.\label{eq:reparam3}
\end{align}
Meanwhile, the parameterized $T$ operators, in cartesian form, are:
\begin{align}
    T_0&=\begin{pmatrix}
        (T_{xx}+T_{yy})/2 & 0 & 0\\
        0 & (T_{xx}+T_{yy})/2 & 0\\
        0 & 0 & T_{zz}
    \end{pmatrix},\\
    T_\pm&=\begin{pmatrix}{}
        (T_{xx}-T_{yy})/4 & \mp i(T_{xx}-T_{yy})/4 & T_{xz}/2\\
        \mp i(T_{xx}-T_{yy})/4 & -(T_{xx}-T_{yy})/4 & \pm i T_{xz}/2\\
        T_{zx}/2 &  \pm i T_{zx}/2 & 0\\
    \end{pmatrix},
\end{align}
which have the corresponding non-zero rank-2 components in spherical tensor form $(T_\alpha)_q^2$ \footnote{Note that we have used the canonical spherical tensor formulations (\textit{e.g.} Table 5.2 in \cite{brown_rotational_2003}) and do not adopt the $\sqrt{6}$ scaling convention nor the $\text{tr}(T_1)=0$ assumption for describing the dipolar hyperfine interactions in \cite{endo_microwave_1982}}:
\begin{align}
    (T_0)^2_0&=(2T_{zz}-T_{xx}-T_{yy})/\sqrt{6},\label{eq:aniso1}\\
    (T_\pm)^2_\mp &= \mp (T_{xz}+T_{zx})/2,\label{eq:aniso2}\\
    (T_{\pm})^2_{\pm 2}& = (T_{xx}-T_{yy})/2.\label{eq:aniso3}
\end{align}{}By applying a Wigner rotation to the molecule-frame ($q$) spherical components, we obtain
\begin{align}
    \bra N' K'|\mathcal{D}^2_{pq}(\omega) (T_\alpha)_q^2 | N K\ket  
    =\sum_q\big[(-1)^{N'-K'}\notag \\
    \times \sqrt{(2N'+1)(2N+1)}\threeJ{N'}{-K'}{2}{q}{N}{K}(T_\alpha)_q^2\big].
\end{align}
We can see that the anisotropic tensor component in eq. (\ref{eq:aniso3}) will couple states with $\Delta K=2$. In the case of $|K|=1$, this term would couple $K=+1$ and $K=-1$ terms, providing a first-order contribution to the $K$-doubling. Most of the sources of $K$-splitting in the hyperfine structure of symmetric tops arise in this manner.

\subsubsection{Stark and Zeeman Matrix Elements}
The Stark Hamiltonian is $\hat{H}_S = -T^1(\mathbf{d}) \cdot T^1(\mathbf{E})$, where $\mathbf{d}$ is the molecule frame dipole moment and $\mathbf{E}$ is the electric field. Without loss of generality, we set the electric field along the $\hat{z}$ direction in the lab frame.  The matrix elements in the decoupled basis $|N, K, S, J, m_J\ket|I_M, m_{IM}\ket |\Gamma, I_H, m_{IH}\ket$ are given by
\begin{widetext}
\begin{align}
&  \bra N'K'S'J'm_J';I_M m_{IM};\Gamma I_H m_{IH}|\hat{H}_{S}  |NKSJm_J;I_M m_{IM};\Gamma I_H m_{IH}\ket \notag \\
&= -E\bra N'K'S'J'm_J'|T^1_0(\mathbf{d})  |NKSJm_J \ket \notag \\
&= -E d_0(-1)^{J'-m_J'} \threeJ{J'}{-m_J'}{1}{0}{J}{m_J}(-1)^{J+N'+1+S'} \sqrt{(2J+1)(2J'+1)}\sixJ{N}{J}{S'}{J'}{N'}{1} \notag \\
& \phantom{=} \times (-1)^{N'-K'} \sqrt{(2N'+1)(2N+1)}\threeJ{N'}{-K'}{1}{0}{N}{K}.   
\end{align}
\end{widetext}
Note that the dipole moment is along the molecular axis such that $T^1_0(\mathbf{d}) = d_0$ and that the matrix element is diagonal with respect to the nuclear spins.

The Zeeman Hamiltonian is written as $\hat{H}_z= \hat{H}_{z,I} +  \hat{H}_{z,R}$, with two terms corresponding to the coupling of the nuclear spin and molecular rotation to the external magnetic field. 

The nuclear spin term $\hat{H}_{z,I}$ is given by
\begin{align}
\hat{H}_{z,I} & = -g_N \mu_N\sum_i T^1(\mathbf{I}_i) \cdot T^1(\mathbf{B}) \notag \\
& = -g_N \mu_N \sum_{i,p} (-1)^p T^1_p(\mathbf{I}_i)T^1_{-p}(\mathbf{B})
\end{align}
where we note that we need to sum over all the hydrogen and radium nuclear spins.
Evaluating the matrix element in the decoupled basis for the radium spin, which is diagonal in $|N, K, S, J, m_J\ket|\Gamma, I_H, m_{IH}\ket$:

\begin{align}
    &\bra I_M' m_{IM}'| \hat{H}_{z,I_M} | I_M m_{IM}\ket=\notag\\
    &-g_N \mu_N\sum_{p} (-1)^p \bra I_M', m_{IM}'|T^1_p(\mathbf{I}_M)|I_M, m_{IM}\ket\bra T^1_{-p}(\mathbf{B})\ket,
\end{align}
where the nuclear spin angular momentum matrix element for a single nucleus is, generically, 
\begin{align}
    \bra I', m_I'|T^1_{p}(\mathbf{I})|I, m_I\ket& =(-1)^{I'-m_I'}\threeJ{I'}{-m_I'}{1}{p}{I}{m_I}\notag\\
    &\times \delta_{I', I}\sqrt{I(I+1)(2I+1)}.\label{eq:TI}
\end{align}
For the coupling to hydrogen spins, we evaluate the parameterized $T^1(\mathbf{I}_0)$ matrix elements derived in the previous section, which results in a near-identical expression to eq. (\ref{eq:TI}),
\begin{align}
    &\bra I_M m_{IM}| \hat{H}_{z,I_M} | I_M m_{IM} \ket\notag=\\
    &-g_N \mu_N\sum_{p}\big[ (-1)^p \bra \Gamma', I_H', m_{IH}'|T^1_p(\mathbf{I}_0)|\Gamma, I_H, m_{IH}\ket\notag \\
    &\times \bra T^1_{-p}(\mathbf{B})\ket\big].
\end{align}{}

The rotational Zeeman term \cite{hirota_high-resolution_1985} is:
\begin{align}
    \hat{H}_{z,R} & = g_R \mu_N \,\mathbf{N} \cdot \mathbf{B} \\
    & = \mu_N B_0 \sum_k (-1)^k\sqrt{\frac{2k+1}{3}}\,T^1_0(g^k_r, \mathbf{N})\notag\\
    & = \mu_N B_0 
     \sum_k (-1)^{2k+1}\sqrt{2k+1}\notag\\
     &\times\sum_p\frac{1}{2}\big[T^k_p(g_r) T^1_{-p}(\mathbf{N})+(-1)^k T^1_p(\mathbf{N})T^k_{-p}(g_r)\big]\notag\\
     &\times \threeJ{k}{p}{1}{-p}{1}{0}.
\end{align}
Relativistic \textit{ab initio} values for the rotational g-tensor $g^k_r$ have been computed using the four-component linear response within elimination of small component (LRESC) approach of Aucar et al. \cite{aucar_theoretical_2014} as implemented in the DIRAC19 code \cite{a_s_p_gomes_dirac_nodate}.  Electron correlation is treated at the level of DFT with a B3LYP functional \cite{stephens_ab_1994} and a dyall.v4z basis is used \cite{dyall_relativistic_2009}. See table \ref{tab:params} in main text for specific values.

\subsubsection{Nuclear Spin-Rotation Coupling}
The nuclear spin-rotation coupling is expressed similarly to the electron spin-rotation elements:
\begin{align}
    H_{\text{nsr}}&=\frac{1}{2}\sum_{\alpha, \beta}C_{\alpha, \beta}({N}_{\alpha}{I}_{\beta}+{I}_{\beta}{N}_{\alpha})\\
    &=\frac{1}{2}\sum^2_{k=0}\big[T^k(C)\cdot T^k(\mathbf{N}, \mathbf{I})+T^k(\mathbf{N}, \mathbf{I})\cdot T^k(C) \big]\\
    &=\frac{1}{2}\sum^2_{k=0}\sum^k_p\big[T^k_p(C)T^k_{-p}(\mathbf{N}, \mathbf{I})+T^k_p(\mathbf{N}, \mathbf{I})T^k_{-p}(C) \big],\label{eq:nsr1}
\end{align}
where the product is decomposed as
\begin{align}
    T^k_p(\mathbf{N}, \mathbf{I})=(-1)^p\sqrt{2k+1}\sum_{p_1, p_2}T^1_{p_1}(\mathbf{N})T^1_{p_2}(\mathbf{I})\threeJ{1}{p_1}{1}{p_2}{k}{-p}.
\end{align}
In $^{225}$RaOCH$_3^+$, the nuclear spin rotation interaction is divided into two components: the interaction between rotation and the spin-$1/2$ $^{225}$Ra nucleus, and the interaction between rotation and the hydrogen spins in the methyl group. 

Let us consider the spin-rotation coupling for the single radium spin $\mathbf{I}_M$. We write out eq. (\ref{eq:nsr1}), in the decoupled basis $|N, K, S, J, m_J\ket |I_M, m_{IM}\ket |\Gamma, I_H, m_{IH}\ket$, taking care to sum over a complete set of states between $T(C)$ and $T(\mathbf{N})$,
\begin{widetext}
\begin{align}
    \bra & N' K' S' J' m_J'; I_M' m_{IM}';\Gamma I_H m_{IH}|H_\text{nsr-M}|N K S J m_J; I_M m_{IM}; \Gamma I_H m_{IH}\ket\notag\\
    &=\frac{1}{2}\sum_{k,p}(-1)^p\sqrt{2k+1}\sum_{p_1, p_2}\bigg[\sum_{\eta''}\delta_{J', J}\frac{(-1)^{J'-J''}}{2J'+1}
    \bra N' K' S' J' m_J'|T^k_p(C)| N'' K'' S'' J'' m_J''\ket\notag\\
    &\times\bra N'' K'' S'' J'' m_J''|T^1_{p_1}(\mathbf{N})| N K S J m_J\ket\bra I_M' m_{IM}'|T^1_{p_2}(\mathbf{I}_M)|I_M m_{IM}\ket\threeJ{1}{p_1}{1}{p_2}{k}{p}\notag\\
    &+\sum_{\eta''}\delta_{J', J}\frac{(-1)^{J'-J''}}{2J'+1}\bra N' K' S' J' m_J'|T^1_{p_1}(\mathbf{N})| N'' K'' S'' J'' m_J''\ket\notag\\
    &\times\bra \Gamma' I_M' m_{IM}'|T^1_{p_2}(\mathbf{I}_M)|\Gamma I_M m_{IM}\ket\bra N'' K'' S'' J'' m_J''|T^k_{-p}(C)| N K S J m_J\ket\threeJ{1}{p_1}{1}{p_2}{k}{-p}\bigg].\label{eq:hnsr} 
\end{align}
\end{widetext}
Note that eq. (\ref{eq:hnsr}) is diagonal with respect to the $|\Gamma, I_H, m_{IH}\ket$ spin states.

Evaluating the nuclear spin-rotation and rotational angular momentum matrix elements yields

\begin{align}
    \bra & N' K' S' J' m_J'|T^1_{p}(\mathbf{N})| N K S J m_J\ket=\notag\\
    &\times (-1)^{J'-m_J'}\threeJ{J'}{-m_J'}{1}{p}{J}{m_J}\notag\\
    &\times\delta_{S', S}(-1)^{J+N'+1+S'}\sqrt{(2J'+1)(2J+1)}\notag\\
    &\times\sixJ{N}{J}{S}{J'}{N'}{1}\delta_{N', N}\sqrt{(N(N+1)(2N+1)},
\end{align}

\begin{align}
    \bra & N' K' S' J' m_J'|T^k_{p}(C)| N K S J m_J\ket=\notag\\
    &(-1)^{J'-m_J'}\threeJ{J'}{-m_J'}{k}{p}{J}{m_J}\notag\\
    &\times\delta_{S', S}(-1)^{J+N'+k+S'}\sqrt{(2J'+1)(2J+1)}\sixJ{N}{J}{S}{J'}{N'}{k}\notag\\
    &\times\sum_q\big[(-1)^{N'-K'}\sqrt{(2N'+1)(2N+1)}\notag\\
    &\times\threeJ{N'}{-K'}{k}{q}{N}{K}T_q^k(C)\big],
\end{align}

\noindent where, for an on-axis spin, the non-zero spherical tensor elements for the nuclear spin-rotation coupling are:
\begin{align}
    T_0^0(C)&=-(C_{xx}+C_{yy}+C_{zz})/\sqrt{3},\\
    T_0^2(C)&=(2C_{zz}-C_{xx}-C_{yy})/\sqrt{6},\\
    T_{\pm 2}^2(C)&=(C_{xx}-C_{yy})/2.
\end{align}

The lab-frame matrix element $T^1_p(\mathbf{I}_M)$ can be evaluated as eq. (\ref{eq:TI}) for a single spin. 

Now, we consider the coupling to rotation for the composite hydrogen spins ($\mathbf{I}_H$). We can utilize the parameterized nuclear spin operators and coupling tensors derived earlier. This adds an additional summation over the parameter $\alpha$,
\begin{align}
    H_\text{nsr-H}&=\frac{1}{2}\sum_\alpha\sum^2_{k=0}\sum^k_p\big[T^k_p(C_\alpha)T^k_{-p}(\mathbf{N}, \mathbf{I}_{-\alpha})\notag\\
    &+T^k_p(\mathbf{N}, \mathbf{I}_{-\alpha})T^k_{-p}(C_\alpha)\big].
\end{align}

\noindent We can thus generalize eq. (\ref{eq:hnsr}) to the case of multiple spins,

\begin{widetext}
\begin{align}
     \bra & N' K' S' J' m_J'; \Gamma' I' m_I'|H_\text{nsr-H}|N K S J m_J;\Gamma I m_I\ket=\sum_\alpha\frac{1}{2}\sum_{k,p}(-1)^p\sqrt{2k+1}\notag\\
    &\times\sum_{p_1, p_2}\bigg[\sum_{\eta''}\delta_{J', J}\frac{(-1)^{J'-J''}}{2J'+1}
    \bra N' K' S' J' m_J'|T^k_p(C_\alpha)| N'' K'' S'' J'' m_J''\ket\notag\\
    &\times\bra N'' K'' S'' J'' m_J''|T^1_{p_1}(\mathbf{N})| N K S J m_J\ket\bra\Gamma' I' m_I'|T^1_{p_2}(\mathbf{I}_{-\alpha})|\Gamma, I, m_I\ket\threeJ{1}{p_1}{1}{p_2}{k}{p}\notag\\
    &+\sum_{\eta''}\delta_{J', J}\frac{(-1)^{J'-J''}}{2J'+1}\bra N' K' S' J' m_J'|T^1_{p_1}(\mathbf{N})| N'' K'' S'' J'' m_J''\ket\notag\\
    &\times\bra\Gamma' I' m_I'|T^1_{p_2}(\mathbf{I}_{-\alpha})|\Gamma I m_I\ket\bra N'' K'' S'' J'' m_J''|T^k_{-p}(C_\alpha)| N K S J m_J\ket\threeJ{1}{p_1}{1}{p_2}{k}{-p}\bigg].
\end{align}
\end{widetext}

\noindent where $\bra\Gamma' I' m_I'|T^1_{p}(\mathbf{I}_{\alpha})|\Gamma I m_I\ket$ is evaluated using Wigner-Eckart and the reduced matrix elements listed in eqs. (\ref{eq:reparam1}) -- (\ref{eq:reparam3}). The $T^k_{p}(C_\alpha)$ tensors can be obtained with the appropriate parameterization listed in eqs. (\ref{eq:reparam1}) -- (\ref{eq:reparam2}). 

Relativistic \textit{ab initio} values for the $T(C)$ nuclear spin-rotation tensors are also obtained with a four-component LRESC approach \cite{aucar_theoretical_2012, aucar_theoretical_2013} in the DIRAC19 code \cite{a_s_p_gomes_dirac_nodate} using a dyall.v4z basis set \cite{dyall_relativistic_2009}. Correlation effects are treated at the level of DFT with the local density approximation and popular hybrid functions (B3LYP \cite{stephens_ab_1994}, PBE0 \cite{adamo_toward_1999}). The full results are listed in Table \ref{tab:diracprops}. While validating these methods is challenging, due to the absence of experimental data (particularly for atoms heavier than group 4), benchmarks with diatomic species in refs. \cite{aucar_theoretical_2014, aucar_theoretical_2013} suggests that these predictions are accurate within 10 percent.

The contributions from nuclear spin rotation couplings to the $K$-doubling can be extracted from the anisotropy in the $T(C)$ tensors. As the radium spin-rotation coupling is diagonal in $|\Gamma, I_H, m_{IH}\ket$, the $H_\text{nsr-M}$ Hamiltonian does not couple off-diagonally between $|E_+\ket$ and $|E_-\ket$ hydrogen spin states, and therefore does not contribute to the breaking of $K$-degeneracy. By contrast, the hydrogen spin-rotation coupling does couple $|E_+\ket$ and $|E_-\ket$ hydrogen spin states. The computed anisotropic contribution to $K$-doubling is $|T_{xx}-T_{yy}|\sim 0.3$ kHz, as noted in Table \ref{tab:params} of the main text. 

\subsubsection{Nuclear Spin Dipolar Coupling}

For coupling two generic spins $I_1$ and $I_2$, the matrix element is

\begin{align}
    H_{II}&=\frac{\mu_0\gamma_1\gamma_2\hbar^2}{4\pi}\bigg[\frac{\mathbf{I}_1\cdot\mathbf{I}_2}{r^3}-\frac{3(\mathbf{I}_1\cdot \mathbf{r})(\mathbf{I}_2\cdot \mathbf{r})}{r^5}\bigg]\\
    &=-\sqrt{6}\frac{\mu_0\gamma_1\gamma_2\hbar^2}{4\pi}T^2(C)\cdot T^2(\mathbf{I}_1, \mathbf{I}_2)\\
    &=-\sqrt{6}\frac{\mu_0\gamma_1\gamma_2\hbar^2}{4\pi}\sum_p T^2_p(C) T^2_{-p}(\mathbf{I}_1, \mathbf{I}_2),
\end{align}

\noindent where $\mu_0$ is the vacuum permeability and $\gamma_{1,2}$ are the nuclear gyromagnetic ratios.  $T^2_q(C)$ is the spherical harmonic tensor $\bra C^2_q(\theta, \phi) r^{-3}\ket$, where the polar angles $(\theta, \phi)$ parameterize the vector from $I_1$ to $I_2$. Meanwhile, $T^2(\mathbf{I}_1, \mathbf{I}_2)$ can be decomposed as

\begin{align}
    T^2_p(\mathbf{I}_1, \mathbf{I}_2)=(-1)^p\sqrt{5}\sum_{p_1, p_2}T^1_{p_1}(\mathbf{I}_1) T^1_{p_2}(\mathbf{I}_2)\threeJ{2}{-p}{1}{p_1}{1}{p_2}.
\end{align}

Let us consider the spin-spin interaction between the $^{225}$Ra nucleus and the hydrogen nuclei. We write the matrix elements in the decoupled basis $|N, K, S, J, m_J\ket |I_M, m_{IM}\ket |\Gamma, I_H, m_{IH}\ket$ and then evaluate the spherical harmonic and nuclear spin-spin tensors
\begin{widetext}
\begin{align}
    \bra N' K' S' J' m_J';& I_M' m_{IM}';\Gamma', I_H' m_{IH}'|H_\text{ss}|N K S J m_J; I_M m_{IM};\Gamma I_H m_{IH}\ket\notag\\
    &=-\sqrt{6}\frac{\mu_0\gamma_1\gamma_2\hbar^2}{4\pi}\sum_p\bra N' K' S' J' m_J'| T^2_p(C)| N K S J m_J\ket \notag \\
    &\times \bra I_M' m_{IM}';\Gamma' I_H' m_{IH}'|T^2_{-p}(\mathbf{I}_M, \mathbf{I}_H)| I_M m_{IM};\Gamma I_M m_{IM}\ket\notag\\
    &=-\sqrt{6}\frac{\mu_0\gamma_1\gamma_2\hbar^2}{4\pi}\sum_\alpha\sum_p\bra N' K' S' J' m_J'| T^2_p(\mathbf{C_\alpha})| N K S J m_J\ket \notag \\
    &\times \bra I_M' m_{IM}';\Gamma' I_H' m_{IH}'|T^2_{-p}(\mathbf{I}_M, \mathbf{I}_{-\alpha})| I_M m_{IM};\Gamma I_M m_{IM}\ket,
\end{align}
\begin{align}
    \bra N' K' S' J' m_J'&| T^2_p(C_\alpha)| N K S J m_J\ket\notag\\
    =&(-1)^{J'-m_J'}\threeJ{J'}{-m_J'}{2}{p}{J}{m_J}\delta_{S', S}(-1)^{J+N'+S'+2}\sqrt{(2J'+1)(2J+1)}\sixJ{N}{J}{S}{J'}{N'}{2}\notag\\
    &\times\bra N' K'|\sum_q \mathcal{D}^2_{pq}(\omega)\cdot T^2_q(C_\alpha)| N K\ket\\
    =&(-1)^{J'-m_J'}\threeJ{J'}{-m_J'}{2}{p}{J}{m_J}\delta_{S', S}(-1)^{J+N'+S'+2}\sqrt{(2J'+1)(2J+1)}\sixJ{N}{J}{S}{J'}{N'}{2}\notag\\
    &\times\sum_{q}(-1)^{N'-K'}\sqrt{(2N'+1)(2N+1)}\threeJ{N'}{-K'}{2}{q}{N}{K}T^2_q(C_\alpha)\\
    \bra I_M' m_{IM}';\Gamma' I_H' m_{IH}'&|T^2_p(\mathbf{I}_M, \mathbf{I}_{\alpha})|I_M m_{IM};\Gamma I_H m_{IH}\ket\notag \\
    =&(-1)^p\sqrt{5}\sum_{p_1, p_2}\bra I_M' m_{IM}'|T^1_{p_1}(\mathbf{I}_M)|I_M m_{IM}\ket\bra\Gamma' I_H' m_{IH}'|T^1_{p_2}(\mathbf{I}_{\alpha})|\Gamma I_H m_{IH}\ket\threeJ{2}{-p}{1}{p_1}{1}{p_2}.
\end{align}
\end{widetext}

The molecule-frame $T^2_q(C_\alpha)$ reduced matrix element can be evaluated using the appropriate sums of spherical harmonics that parameterize the vector between the $i$th hydrogen and the radium atom,

\begin{align}
    T^2_q(\mathbf{C}_0)&=\sum_{i=1}^3\sqrt{\frac{4\pi}{5}}Y^2_q(\theta_i, \phi_i)r^{-3},\\
    T^2_q(\mathbf{C}_\pm)&=\sqrt{\frac{4\pi}{5}}\big[Y^2_q(\theta_1, \phi_1)+e^{\pm 2\pi i/3}Y^2_q(\theta_2, \phi_2)\notag\\
    &+e^{\mp 2\pi i/3}Y^2_q(\theta_3, \phi_3)\big]r^{-3}.
\end{align}
By convention \cite{endo_microwave_1982}, the Cartesian form of the single-hydrogen dipolar tensor $T_1$ (of the form eq. (\ref{eq:t1})) is taken to be traceless $\text{tr}[T]=0$ and symmetric, which yields the following simplifications from eqs. (\ref{eq:aniso1})--(\ref{eq:aniso2}), which are invoked in table \ref{tab:params} in the main text:
\begin{align}
    \alpha_{dip}\cdot T^2_0(\mathbf{C}_0)&=3T_{zz}/\sqrt{6},\\
    \alpha_{dip}\cdot T^2_{\pm}(\mathbf{C}_\mp)&=\mp T_{xz},\\
    \alpha_{dip}\cdot T^2_{\pm 2}(\mathbf{C}_\pm)&= (T_{xx}-T_{yy})/2.
\end{align}
where we define a scaling parameter $\alpha_{dip}=-\sqrt{6}\mu_0\gamma_{\text{H}}\gamma_{\text{Ra}}\hbar^2/4\pi$.

The radium-hydrogen spin-spin interaction couples the opposite symmetry states $|E_{+}\ket$ and $|E_{-}\ket$. This leads to $|T_{xx}-T_{yy}|\sim 0.4$ kHz $K$-doubling terms which couple the $|N, K, S, J, m_J\ket|I_M, m_{IM}\ket|I_H, m_{IH}\ket$ and  $|N, -K, S, J, m_J\ket|I_M, m_{IM}\ket|I_H, m_{IH}\ket$ states. The spin-spin interaction between the hydrogen spins themselves are only non-zero between the ``ortho'' stretched spin states of $A$ character. This term can therefore be ignored in the $|N, |K|=1\ket$ manifold.

\section{Schiff moment}

\subsection{Effective molecular sensitivity}

The effective Schiff moment sensitivity is proportional to $\bra  \mathbf{I}_M \cdot \mathbf{\hat{n}} \ket$, where $I_M$ is the metal spin and $\hat{n}$ is the internuclear axis \cite{flambaum_enhanced_2019}. In the decoupled basis, this is written as: 
\begin{widetext}
\begin{align}
    \bra & N', K', S', J', m_J';I_M', m_{IM}';I_H', m_{IH}' |\mathbf{I}_M \cdot \mathbf{\hat{n}}|N, K, S, J, m_J;I_M, m_{IM};I_H, m_{IH}\ket\notag\\
    &= \delta_{I_H', I_H}\delta_{m_{IH}', m_{IH}}\bra I_M', m_{IM}'|\mathbf{I}_M|I_M, m_{IM}\ket \bra N', K', S', J', m_J'|\hat{n}|N, K, S, J, m_J\ket \notag\\
    &= \delta_{I_H', I_H}\delta_{m_{IH}', m_{IH}}\bra I_M', m_{IM}'|\mathbf{I}_M|I_M, m_{IM}\ket(-1)^{J'-m_J'}\threeJ{J'}{-m_J}{1}{0}{J}{m_J}\notag \\
    &\phantom{=}\times\delta_{S', S}(-1)^{J+N'+S'+1}\sqrt{(2J'+1)(2J+1)}\sixJ{N}{J}{S'}{J'}{N'}{1}\bra N'|\mathbf{\hat{n}}|N \ket \notag\\
    &= \delta_{I_H', I_H}\delta_{m_{IH}', m_{IH}}(-1)^{I_M'-m_{IM}'}\threeJ{I_M'}{-m_{IM}'}{1}{0}{I_M}{m_{IM}}\delta_{I_M', I_M}\sqrt{I_M(I_M+1)(2I_M+1)}\notag \\
    &\phantom{=}\times(-1)^{J'-m_J'}\threeJ{J'}{-m_J}{1}{0}{J}{m_J}\delta_{S', S}(-1)^{J+N'+S'+1}\sqrt{(2J'+1)(2J+1)}\sixJ{N}{J}{S'}{J'}{N'}{1}\notag\\
    &\phantom{=}\times(-1)^{N'-K'}\sqrt{(2N'+1)(2N+1)}\threeJ{N'}{-K'}{1}{0}{N}{K}
\end{align}{}
\end{widetext}
In zero field, there is no defined orientation of the molecule, and the Schiff moment sensitivity is zero. In the decoupled limit ($\gtrsim 1$ V/cm), the value of $\bra  \mathbf{I}_M \cdot \mathbf{\hat{n}} \ket$ is $K \cdot m_N\cdot m_{IM}\cdot 1/2$ for the $N=1$ and $|K|=1$ manifold. The $1/2$ factor arises from the $K/N(N+1)$ pre-factor to the Stark energy. The states with ``stretched" sensitivity therefore take on the values $\bra  \mathbf{I}_M \cdot \mathbf{\hat{n}} \ket=\pm 1/4$.
\subsection{BSM Sensitivity}
The frequency sensitivity of a spin-precession measurement on a molecule or atom with coherence time $\tau$ and repetition $n$ is expressed as $\delta \omega = [\tau \sqrt{n}]^{-1}$. In the main text, we assume a single trapped $^{225}$RaOCH$_3^+$ ion with 5 sec coherence time, which provides a frequency sensitivity of $\delta\omega=7.5$ mrad s$^{-1}/\sqrt{\text{hour}}$.

Dobaczewski and Engel calculated $|\mathbf{S}(^{225}$Ra$)|$ in the framework of $T,P$-violating interactions between nucleons mediated by a pion using a variety of Skyrme energy functionals \cite{dobaczewski_nuclear_2005}. The average of their results, which was expressed in terms of $\pi N N$ vertices, was re-parameterized by ref. \cite{flambaum_enhanced_2019} in terms of QCD $\bar{\theta}$ and the quark-chromo EDMs $\tilde{d}_u$ and $\tilde{d}_d$:
\begin{align}
    |\mathbf{S}(^{225}\text{Ra})|&= 1.0 \bar{\theta} \text{ \textit{e} fm}^3\\
    &=10^4(0.50\tilde{d}_u-0.54\tilde{d}_d)\text{ \textit{e} fm}^2
\end{align}

The energy shift resulting from a Schiff moment $\mathbf{S}$ and a coupling constant $W_s$ is
$H_\text{sm}=W_s\, (\mathbf{I}_M\cdot \mathbf{\hat{n}})\,|\mathbf{S}|/|\mathbf{I}|$. For a differential measurement performed between states of opposite effective sensitivity ($\bra  \mathbf{I}_M \cdot \mathbf{\hat{n}} \ket=\pm 1/4$), we arrive at
\begin{align}
    \delta\bar{\theta}=\frac{2\hbar\delta\omega}{W_S(S/\bar{\theta})}
\end{align}
where $S/\bar{\theta}=1.0\, e \text{ fm}^3$ and the factor of two arises from $\Delta[\bra  \mathbf{I}_M \cdot \mathbf{\hat{n}} \ket]=1/2$. Given the estimate of $W_s($RaOCH$_3^+)\approx 30,000$ a.u. \cite{flambaum_enhanced_2019} (where a.u. $\equiv e/4\pi\epsilon_0a^4_0$), we have $\delta\bar{\theta}/\delta\omega=24\times 10^{-8}$ s. Two weeks of data taking would thus result in the model-dependent sensitivity of $\delta\bar{\theta}<\times 10^{-10}$.

This discussion only considers in detail the limit on $\bar{\theta}$ as an example, though the other quantities to which Schiff moments are sensitive are also of interest.  Due to the many possible hadronic sources, measurements in different systems are needed to obtain robust bounds \cite{chupp_electric_2015}

\section{Spin Zero Isotopologues}
\begin{figure}
	\includegraphics[width=.9\columnwidth]{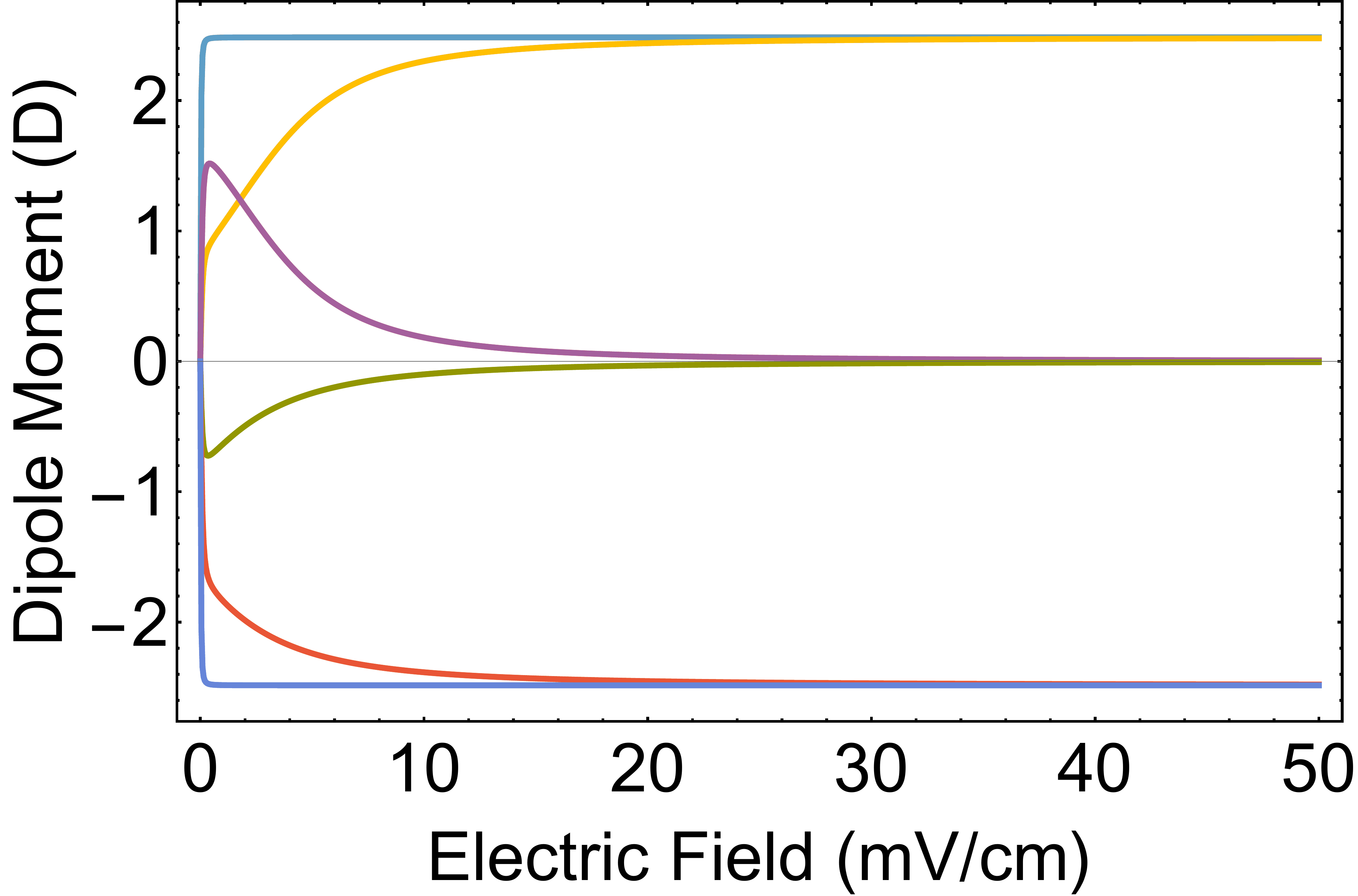}
	\includegraphics[width=.9\columnwidth]{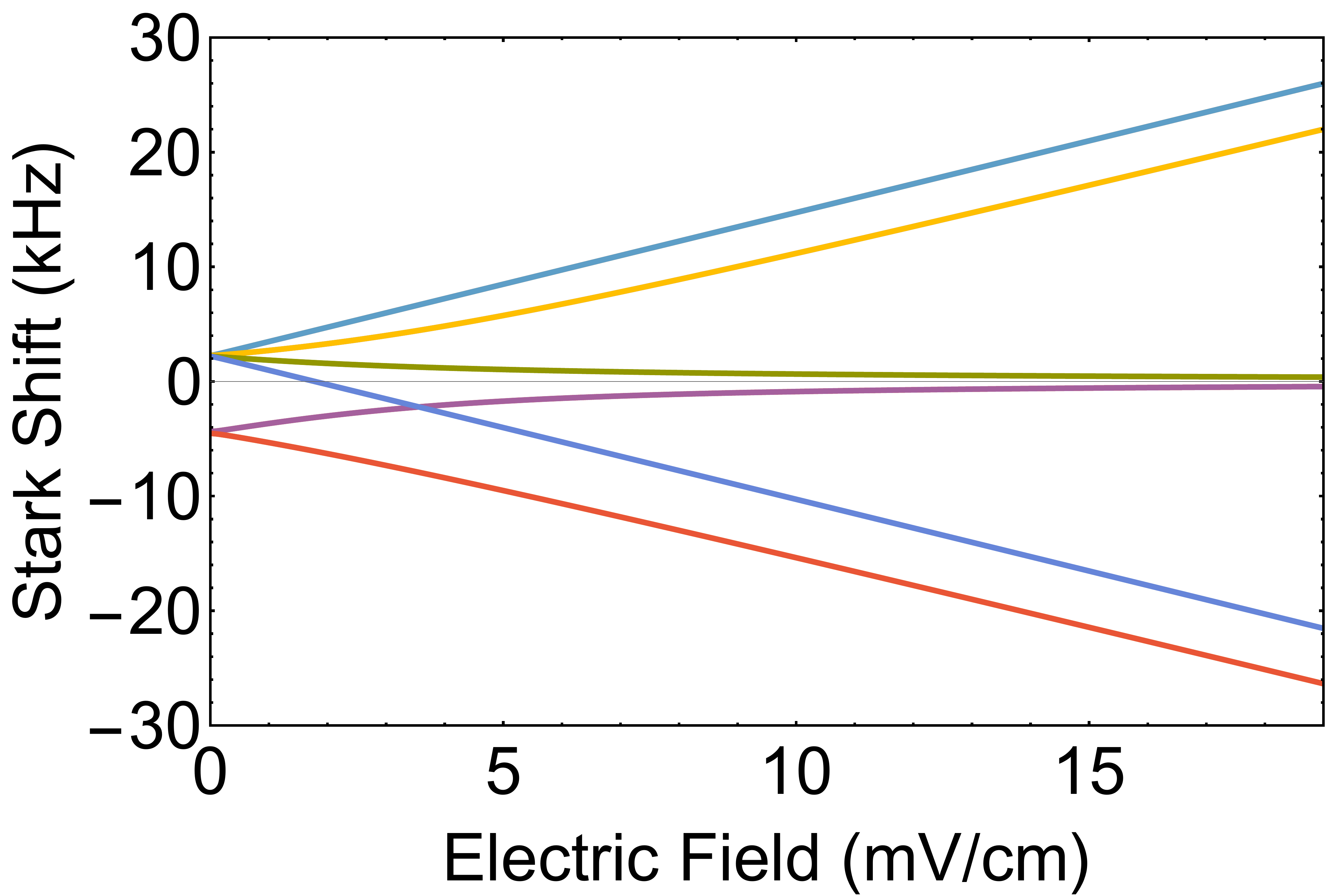}
	\caption{Lab frame dipole moment (top) and Stark shifts (bottom) of hyperfine states of the $|N=1,|K|=1\ket$ manifold in the fully polarized limit for $^{226}$RaOCH$_3^+$.  High, low, and no field seekers correspond to states with negative, positive, and zero dipole moment ($K\times m_N= +1$, $0$, and $-1$).	The jumps indicate avoided crossings between the Stark states.}
	\label{fig:226stark}
\end{figure}

\begin{figure}
	\includegraphics[width=\columnwidth]{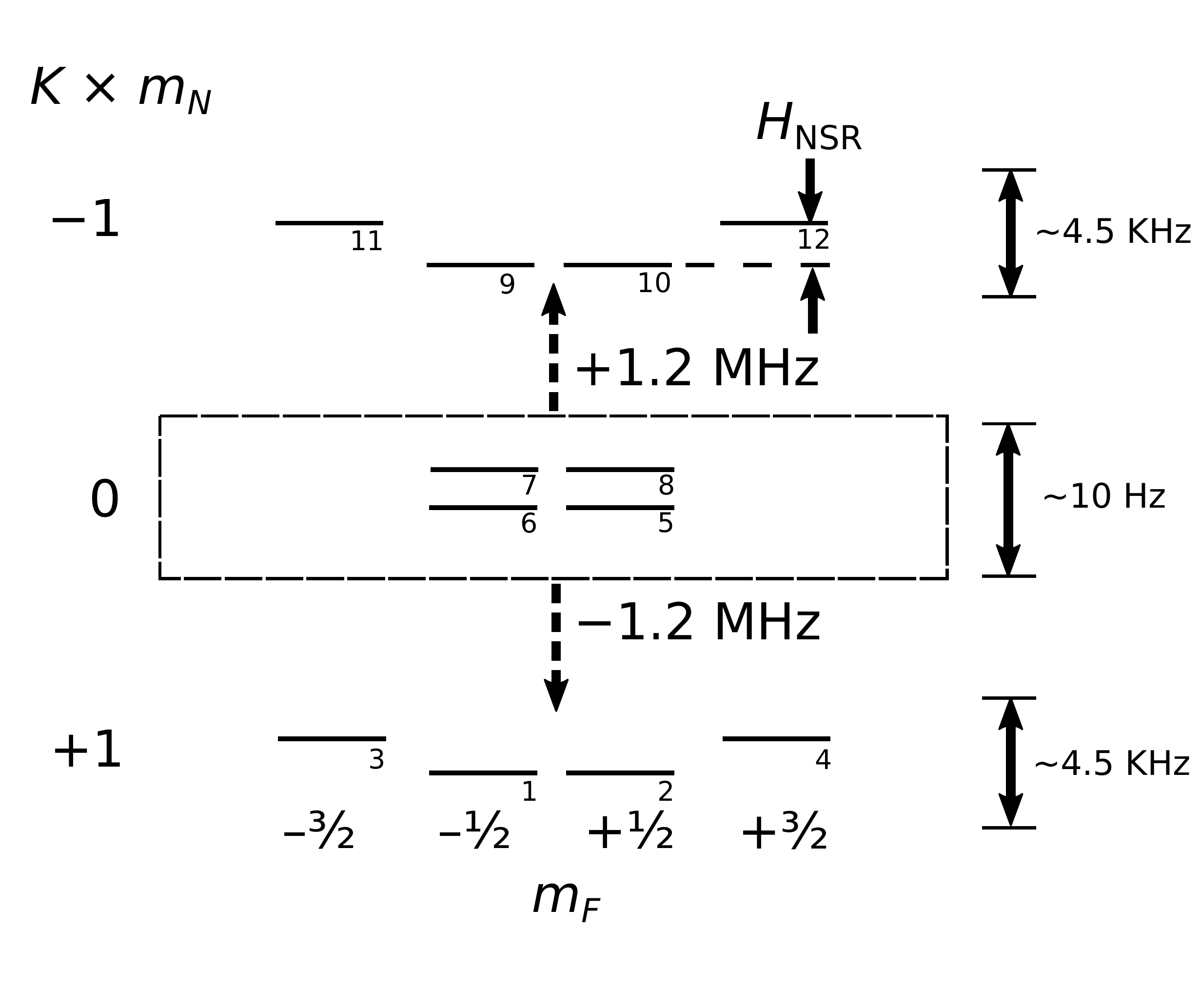}
	\caption{Level structure for the 12 hyperfine states of $^{226}$RaOCH$_3^+$ in the $|N=1,|K|=1\ket$ manifold in the decoupled regime ($1$ V/cm), grouped by the projection of total angular momentum $m_F=m_N+m_{IH}$ and their Stark manifold ($K\times m_N$). The states are numbered in order of energy.}
	\label{fig:226level}
\end{figure}

We briefly consider the hyperfine structure for $^{226}$RaOCH$_3^+$ in the $|N=1, |K|=1\ket$ manifold, which yields 12 unique levels.  This is the isotopologue recently produced in an ion trap \cite{fan_optical_2020}, and could serve as a valuable platform for spectroscopy and development.  As the $^{226}$Ra nucleus is spin-0, the only hyperfine term to consider is the spin-rotation interaction between the ortho-hydrogen state and the molecular rotation, and the hyperfine basis reduces to $|N, K, S, J, m_J\ket|\Gamma, I_H, m_{IH}\ket$ with the radium spin terms omitted. Fig. \ref{fig:226stark} shows the Stark shifts and dipole moments for the molecule up to the fully polarized limit, and fig. \ref{fig:226level} shows the level structure grouped by the projection of total angular momentum ($m_F=m_N+m_{IH}$) and their Stark manifold ($K\times m_N$).

\begin{table*}[h]
    \centering
    \begin{ruledtabular}
    \begin{tabular}{lrrrr}
        \textbf{Method:} & \shortstack[r]{ LRESC(B3LYP)} & \shortstack[r]{LRESC(PBE0)} & \shortstack[r]{LRESC(LDA)} & \shortstack[r]{LRESC(DHF)} \\
        \textbf{Basis}: & \shortstack[r]{dyall.v4z} & \shortstack[r]{dyall.v4z} & \shortstack[r]{dyall.v4z} &
        \shortstack[r]{dyall.v4z}\\
        \hline
        NSR (H) \\
        \hline
        $T(C_\text{nsr})_{zz}$ & 15.27503794 & 15.23988653 & 15.05936414 & 15.54351836 \\
        $T(C_\text{nsr})_{yy}$ & 0.29172044 & 0.29285025 & 0.27717575 & 0.31026183 \\
        $T(C_\text{nsr})_{xx}$ & -0.00937583 & -0.00725048 & -0.02242052 & 0.01481825 \\
        $T(C_\text{nsr})_{zy}$ &  -0.00000000 & -0.00000000 & -0.00000000 & -0.00000000 \\
        $T(C_\text{nsr})_{yz}$ & -0.00000012 & -0.00000012 & -0.00000010 & -0.00000014 \\
        $T(C_\text{nsr})_{yx}$ & -0.00000005 & -0.00000005 & -0.00000005 & -0.00000005 \\
        $T(C_\text{nsr})_{xy}$ & 0.00000005 & 0.00000005 & 0.00000005 & 0.00000005 \\
        $T(C_\text{nsr})_{xz}$ & 253.81827070 & 253.90232417 & 254.10917668 & 253.71558497 \\
        $T(C_\text{nsr})_{zx}$ & -3.38051883 & -3.37994662 & -3.37277269 & -3.38404825  \\
        \hline
        NSR (Ra) \\
        \hline
        $T(C_\text{nsr})_{xx}$ & 3.67255479 & 3.52587436 & 4.86603582 & 1.71006450 \\
        $T(C_\text{nsr})_{yy}$ & 2.41881565 & 2.49385127 & 3.24531289 & 1.35943566 \\
        $T(C_\text{nsr})_{xx}$ & 2.41881768 & 2.49385648 & 3.24531347 & 1.35945538 \\
        $T(C_\text{nsr})_{zy}$ &  -0.00000003 & -0.00000004 & -0.00000005  & -0.00000002 \\
        $T(C_\text{nsr})_{yz}$ & -0.00000275 & -0.00000284 & -0.00000369 & -0.00000155 \\
        $T(C_\text{nsr})_{yx}$ & 0.00000000 & -0.00000000 & -0.00000000 & 0.00000000 \\
        $T(C_\text{nsr})_{xy}$ & 0.00000000& -0.00000000 & 0.00000000 & -0.00000000 \\
        $T(C_\text{nsr})_{xz}$ & -0.00000158 & -0.00002529 & 0.00000641 & -0.00013274 \\
        $T(C_\text{nsr})_{zx}$ & 0.00000009 & -0.00000060 & -0.00000024 & 0.00000726  \\
    \end{tabular}
    \end{ruledtabular}
    \caption{Nuclear spin-rotation constants computed with optimized geometries from table \ref{tab:params}. Units in kHz.}
    \label{tab:diracprops}
\end{table*}

\begin{sidewaystable*}[]
    \centering
    \begin{ruledtabular}
    \begin{tabular}{c|c|c|c}
         State & Energy & Mixing & Schiff Sensitivity \\
         \hline
         1 & $-1.2545$ MHz & $\big[|-1, -1\ket|+1/2\ket|E_+,+1/2\ket-|+1, +1\ket|-1/2\ket|E_-,-1/2\ket\big]/\sqrt{2}$ &  $0$\\
         2 & $-1.2545$ MHz & $\big[|-1, -1\ket|+1/2\ket|E_+,+1/2\ket+|+1, +1\ket|-1/2\ket|E_-,-1/2\ket\big]/\sqrt{2}$ &  $0$\\
         3 & $-1.2515$ MHz & $|-1, -1\ket|-1/2\ket|E_+,+1/2\ket$ &  $-1/4$\\
         4 & $-1.2515$ MHz & $|+1, +1\ket|+1/2\ket|E_-,-1/2\ket$ &  $+1/4$\\
         5 & $-1.24926$ MHz & $|-1, -1\ket|+1/2\ket|E_+,-1/2\ket$ &  $+1/4$\\
         6 & $-1.24926$ MHz & $|+1, +1\ket|-1/2\ket|E_-,+1/2\ket$ &  $-1/4$\\
         7 & $-1.24781$ MHz & $|+1, +1\ket|+1/2\ket|E_-,+1/2\ket$ &  $+1/4$\\
         8 & $-1.24781$ MHz & $|-1, -1\ket|-1/2\ket|E_+,-1/2\ket$ &  $-1/4$\\
         9 & $-932.31$ Hz & $\big[|-1, 0\ket|+1/2\ket|E_+,-1/2\ket+|-1, 0\ket|-1/2\ket|E_+,+1/2\ket-|+1, 0\ket|+1/2\ket|E_-,-1/2\ket-|+1, 0\ket|-1/2\ket|E_-,+1/2\ket\big]/2$ &  $0$\\
         10 & $-932.287$ Hz & $\big[|-1, 0\ket|+1/2\ket|E_+,-1/2\ket-|-1, 0\ket|-1/2\ket|E_+,+1/2\ket-|+1, 0\ket|+1/2\ket|E_-,-1/2\ket+|+1, 0\ket|-1/2\ket|E_-,+1/2\ket\big]/2$ &  $0$\\
         11 & $-619.612$ Hz & $\big[|-1, 0\ket|+1/2\ket|E_+,-1/2\ket+|-1, 0\ket|-1/2\ket|E_+,+1/2\ket+|+1, 0\ket|+1/2\ket|E_-,-1/2\ket+|+1, 0\ket|-1/2\ket|E_-,+1/2\ket\big]/2$ & $0$ \\
         12 & $-619.584$ Hz & $\big[|-1, 0\ket|+1/2\ket|E_+,-1/2\ket-|-1, 0\ket|-1/2\ket|E_+,+1/2\ket+|+1, 0\ket|+1/2\ket|E_-,-1/2\ket-|+1, 0\ket|-1/2\ket|E_-,+1/2\ket\big]/2$ &  $0$\\
         13 & $619.367$ Hz & $\big[|-1, 0\ket|+1/2\ket|E_+,+1/2\ket+|+1, 0\ket|+1/2\ket|E_-,+1/2\ket\big]/\sqrt{2}$ & $0$ \\
         14 & $619.367$ Hz & $\big[|-1, 0\ket|-1/2\ket|E_+,-1/2\ket+|+1, 0\ket|-1/2\ket|E_-,-1/2\ket\big]/\sqrt{2}$ &  $0$\\
         15 & $932.466$ Hz & $\big[|-1, 0\ket|+1/2\ket|E_+,+1/2\ket-|+1, 0\ket|+1/2\ket|E_-,+1/2\ket\big]/\sqrt{2}$ & $0$ \\
         16 & $932.466$ Hz & $\big[|-1, 0\ket|-1/2\ket|E_+,-1/2\ket-|+1, 0\ket|-1/2\ket|E_-,-1/2\ket\big]/\sqrt{2}$ & $0$ \\
         17 & $1.24705$ MHz & $\big[|-1, +1\ket|-1/2\ket|E_+,-1/2\ket+|+1, -1\ket|+1/2\ket|E_-,+1/2\ket\big]/\sqrt{2}$ & $0$ \\
         18 & $1.24705$ MHz & $\big[|-1, +1\ket|-1/2\ket|E_+,-1/2\ket-|+1, -1\ket|+1/2\ket|E_-,+1/2\ket\big]/\sqrt{2}$ &  $0$\\
         19 & $1.25004$ MHz & $|+1, -1\ket|-1/2\ket|E_-,+1/2\ket$ & $+1/4$ \\
         20 & $1.25004$ MHz & $|-1, +1\ket|+1/2\ket|E_+,-1/2\ket$ & $-1/4$ \\
         21 & $1.25227$ MHz & $|-1, -1\ket|-1/2\ket|E_+,+1/2\ket$ &  $-1/4$\\
         22 & $1.25227$ MHz & $|+1, -1\ket|+1/2\ket|E_-,-1/2\ket$ &  $+1/4$\\
         23 & $1.25371$ MHz & $|+1, -1\ket|-1/2\ket|E_-,-1/2\ket$ & $+1/4$ \\
         24 & $1.25371$ MHz & $|-1, +1\ket|+1/2\ket|E_+,+1/2\ket$ & $-1/4$
    \end{tabular}
    \end{ruledtabular}
    \caption{Energy, mixings, and Schiff sensitivity of hyperfine states of $^{225}$RaOCH$_3^+$ in the $|N=1, |K|=1\ket$ manifold in the high-field regime (1 V/cm). The mixings are written in the decoupled basis $|K, m_{N}\ket|m_{IM}\ket|\Gamma,m_{IH}\ket$. The quantum numbers $N=1$, $I_M=1/2$, and $I_H=1/2$ are omitted for brevity.}
    \label{tab:states_highfield}
\end{sidewaystable*}

\end{document}